\documentclass{article}
\usepackage{xcolor}
\usepackage{graphicx}
\usepackage{amsmath}

\newcommand{\ave}[1]{\langle #1\rangle}

\makeindex 
\begin{document} 

\setcounter{page}{1}

\title{Evolutionary models for simple biosystems}
\author{Franco Bagnoli}
\maketitle
\tableofcontents

\section*{Abstract}

The concept of evolutionary development of structures constituted a \emph{real} revolution in biology: it was possible to understand how the very complex structures of life can arise in an out-of-equilibrium system. The investigation of such systems has shown that indeed, systems under a flux of energy or matter can self-organize into complex patterns, think for instance to Rayleigh-Bernard convection, Liesegang rings, patterns formed by granular systems under shear. Following this line, one could characterize life as a state of matter, characterized by the slow, continuous process that we call evolution. In this paper we try to identify the organizational level of life, that spans several orders of magnitude from the elementary constituents to whole ecosystems. 

Although similar structures can be found in other contexts like ideas (memes) in neural systems and self-replicating elements (computer viruses, worms, etc.) in computer systems, we shall concentrate on biological evolutionary structure, and try to put into evidence the role and the emergence of network structure in such systems.

\section{Introduction}

The study of evolution has been largely aided by theoretical and computer models. First, it is hard to perform experiments in evolution, although some of them has been carried out. As a consequence, one cannot observe the dynamics of an evolutionary phenomenon, but only a (partial) snapshot of some phases. In general, there are many dynamics compatible with these observations, and therefore it is not easy to decide among the possible ``sources'' of an observed behavior. The decision is often quantitative: many hypotheses can in principle originate an observed evolutionary feature, and one has not only to decide which one is compatible with other characteristics, but also which is the more \emph{robust}, \emph{i.e.}, which requires less parameters  (that in general are poorly known)  and less tuning.

An unique characteristic of evolution is its historical character. The genotypic space is so large and fiddled with ``holes'' (corresponding to non-viable phenotypes), there is no chance to observe twice the same genotype in evolution. Similar phenotypes may emerge (convergent evolution). Due to the autocatalytic (either you eat or you are eaten) character of ecosystems, once a branch of a bifurcation is taken, the other alternative is forever forbidden (this generally corresponds to extinctions).

It is difficult to carry out a completely theoretical approach. Most of people interested in theoretical results want to apply their results to the interpretation of real biological systems, secondly, in some 3 billion years, evolution has developed such an intricate variety of ``case studies'' of applied evolution that we still have a lot of poorly understood examples to be studied. And finally, the difficulty in performing experiments, discourages the study of effects that \emph{could} be compatible with evolution but have never been realized on the earth.

The theoretical approach allows to recognize that evolutionary dynamics applies not only to biological system. There are at least other two environments that support (or may support) an evolutionary dynamics. The first one is the human mind, and in this case the replicators are concepts, ideas and myths that can be tracked in human culture. Indeed, ideas may propagate from individual to individual, can mutate and are selected for their ``infectivity'' and their fitness to the cultural corpus. Many cultural traits like religions proved quite capable of propagating, in spite of the load and the detrimental effects they have on their hosts. The term \emph{meme} (sounding like gene and resembling memory) has been coined for this replicator~\cite{memetic}. 
A particularly important aspect of cultural evolution concerns languages, and in particular language competition~\cite{language}. We have not enough space to exanime these aspects here.

The second example is given by the giant computer network, Internet. Computer viruses and worms are examples of replicating objects that can mutate and are selected by their ability of escaping antiviral software and infecting other computers. With the increasing power of computers and their pervasivity, it is expected that such ``life'' forms will be much more common. Computer life forms and memes may cooperate, think for instance to hoaxes, e-mail viruses, etc. We shall concentrate here on ``standard'', ``biological'' life.

Some parts of this paper were previously published in Ref.~\cite{Bagnoli:Review}. See also Ref.~\cite{Baake:review}. 

\section{The structure of an evolving system}

The three basic blocks of evolutionary dynamics are replication, mutation and selection. Evolution works on a population of reproducing individuals, often called replicators.

\subsection{Modeling living individuals}

In order to visualize the problem, let us consider a pool of bacteria, growing in a liquid medium in a well-stirred container, so that we do not have to be concerned with spatial structures. We have chosen bacteria  because they are able to synthesize most of the compounds they need from simple chemical sources, and appear to have a simpler structure than eukaryotes (cells with nucleus) and archea. The ultimate reason for this simplicity is evolution: bacteria have ``chosen'' to exploit the speed of replication in spite of complexity, and optimized accordingly their working machinery.
Viruses have progressed a lot in the direction of speed, but they need a much more structured background: the presence of a chemical machinery assembled by other living cells.

Bacteria absorb nutrients (energy), amino acids and other chemicals (building blocks) from the medium, and uses them to increase their size, up  to a point in which they divide in two (or sprout some buddies). These tasks are  carried out by a biochemical network of proteins. Proteins act as enzymes, transforming chemical elements, as structural elements and  have also a regulatory functions either by directly activating or inactivating other proteins, or by promoting or blocking the production of other or same proteins. In fact, in a growing bacteria, there is continuously the need of producing new proteins and all the other constituents of the body.

A protein is synthesized as an one-dimensional chain of amino acid. The tri-dimensional shape of a protein, and thus its chemical function, is defined (at least in many simpler cases) by its one-dimensional sequence, that folds and assumes its working conformation by itself.\footnote{Many larger complexes are formed by more that just one protein, and some protein need the help of
other enzymes to stabilize the three-dimensional shape with covalent bonds or to add metals ions, sugar chains and other elements.} There are only 20 amino acids used by living beings. The one-dimensional sequence of a protein is \emph{coded} into the one-dimensional sequence of basis (gene) in a chromosome. Many bacteria have one large chromosome, plus a variable number of smaller ones (plasmids). A chromosome is a double helix of DNA, that can be view (in computer terms) as a sequence of symbols from a four-letter alphabet (ATCG). The translation of a gene into a protein takes place in several steps. First the gene is transcribed into an intermediate form, using RNA. In the RNA alphabet, thymine (T) is replaced by uracil (U). This messenger RNA (mRNA) in eukaryotes are further processed by splicing pieces (introns), but the basic working is similar.\footnote{Actually, in archea, too, there are introns, and also in bacteria self-splicing portions of RNA are present.}

Messenger RNA is then translated into a protein by a large complex of proteins and RNA, called ribosome. A ribosome acts as a catalyst, by allowing other small RNA-amino acid complexes called tRNA to bind to mRNA. This binding is rather specific. A tRNA present a triplet of basis that must complement (with some tolerance) to the triplet (called codon) in the specific region of mRNA that is processed by the ribosome. The tRNA carries an amino acid that is specific to its anti-codon (this pairing is performed by other enzymes). Thus, by this large biochemical network, we have the production of a protein following the information stored in a gene.

When a cell has grown sufficiently, it divides. In order to perform this step, also the DNA has to be replicated, using again other proteins and RNA fragments. Actually, in  growing bacteria, both processes (production of body material and DNA copying) are performed in parallel, so that some part of the chromosome can be present in multiple copies. The splitting of a cell into two can also be seen as balance between the exponential growth of the body and the (almost) linear copying of the DNA. All life long, but especially during the copy phase, \emph{mutations} may occur, changing some basis, inserting or deleting others, etc. Mutations are essential to generate diversity, but, as we shall see, they need to occur quite rarely. Clearly, in multi-cellular organisms, only mutations that occur in the germinal line are transmitted to the progeny. 

The translation of a gene depends by other proteins: they can block the translation by binding themselves to DNA, or by altering the DNA itself (methylation). Other proteins are also needed to promote the transcription of a gene. Thus, from the proteins' point of view, a gene is just a way of producing other proteins, under the influence of the specific pool of proteins that are working together into the organism. 

Not all genes are translated into proteins. For instance, tRNA genes are only transcribed into RNA. Other portions of DNA are never transcribed, but can alter the biochemistry of the cell. For instance, specific sequences may inhibit the expression of genes by bending the DNA, or preventing the binding of promotor proteins. Other DNA sequences may interfere with the process of DNA copying, for instance by increasing the number of repeats of themselves (microsatellites, ALU). Other sequences may \emph{jump} from a position of the chromosome to another (transposons), eventually carrying some portion of DNA with itself. By this mechanism they can inactivate, activate or merge genes. 

Bacteria can also alter their genetic contents by capturing or exchanging pieces of DNA (plasmids), a process that resemble the sexual reproduction of some eukaryotes. 

Finally, retroviruses can alter the genome of the host by inserting their own code (or a precursor of it in RNA retroviruses). This process, that \emph{reverts} the usual information path from DNA to RNA and proteins (the so-called ``central dogma'' of biology), is reputed to be one of the most powerful mechanisms of genetic transformations. Transposons are probably the relics of viruses, and a substantial part of some eukaryotic genome can be recognised as inactivated viruses.

The last important element is selection. In our example, bacteria living in a test tube will proliferate, duplicate and probably differentiate. Sooner or later, they will consume all the nutrients in the solution. A few of them can survive sometime by ``eating'' others, but clearly, in a closed environment, the last fate is the extinction of all living forms.\footnote{Actually, this can take a long time, since many bacteria are able to ``freeze'' themselves in form of spores, that can survive for a long time in rather harsh environments.} Life (replicators)  needs an open environment, with fluxes of energy and matter. We can simulate this in a lab, by picking a drop of the culture, and putting it into a fresh medium. In this way, the survival from a test tube to another is a random process, with a vanishing probability from the point of view of individuals. However, when viewed by genes, there are many individuals that share the same genetic information, while other ones may have a mutated version. The genetic selection  favors the genes that are present in a larger number of copies, and have therefore more chances to survive. The growth of bacteria just after being inoculated into a fresh medium is exponential, so those genomes that allow a faster replication of individuals will be present in more copies. This is the winning strategy of most  bacteria. The speed is not the only quantity that can be \emph{optimized}. One could replicate more slowly, but produce some toxic chemical to which it is immune. Or, if the refreshing occurs lately, one could prefer to develop ways of surviving starvation. Finally, if there are oscillations of temperature, acidity, or in the presence of predators, a better strategy could be that of developing ways to overcome these accidents. All these additional instruments need more proteins, more DNA coding, and therefore a slower replications speed. How can the optimum compromise between speed and complexity be reached?

The main idea of evolution is that this is a self-organized process. Mutations produce variety, and selection prunes it. We have to stress that selection is not an absolute limit, it depends on the rest of the environment. The probability of a given genome to pass to next ``flask'' generation (the \emph{fitness}) does not depend on how fast in absolute it replicates, but on how fast it replicates in comparisons with the other genotypes present in the environment. And, as soon as the ``best'' genotype, possibly recently arisen by a mutation, is picked up and colonizes a new flask, its comparative advantage vanishes. Due to selection, copies of it become more numerous, and the competition becomes harder. This effect, named ``red queen'' is the main motor of evolution.

But let us dig more on this subject. The way in which an individual if perceived by others, or influences the environment is called its ``phenotype''. It is a product of proteins (mainly) and therefore of the genes in its genome. It may depend also on the age of the individual, on past experiences, and on the interactions with other genomes (often very similar, as we shall see). A genome is a ``bag of genes'', extending the ``definition'' (not very mathematical) of gene to all genetic elements that have a recognizable persistence along generations, and that influence in some way their phenotype. 

A gene is successful and tends to spread among the population if it confers a selective advantage to the individuals that carry it, and this happens through its expression, \emph{i.e.}, the phenotype. The phenotype is in general a rather complex functions of the genes that constitute it. Genes very rarely control directly some aspect of the phenotype. What a gene does, is to produce a protein, that participates to the intricate biochemical machinery of the cell, and in multicellular organisms to the development of the organism. Variants of the gene (alleles) produces similar proteins in other bodies. The mechanism of blind production of variant plus selection is able to ``optimize'' this assembly of organisms. This optimization is always at short term: the competition (red queen) forbids the development or maintenance of ``accessories'' that may be useful in the future, or that may lead to the development of new functions. Selection tends to prefer quick-and-dirty solutions.\footnote{This effect is particularly evident in the fact that birds living in islands with no predator tend to lose the capacity of flying: a non-flying relative that can invest more energy in egg production tends to be selected, even if when predators reappear it would be useful to have working wings.}
 
We have to consider that the phenotype is often best understood as the result of a pool of genotypes, not as a characteristic of an individual. For instance, an ant or termite nest is the result of the cooperation of several individuals, and of many generations. A multicellular body is the result of the cooperation of many cells. What these cells have in common is the sharing of large portions of their genome (all for multicellular bodies). In other cases (symbionts, larger communities like human societies), cooperation tends to favor assemblies of rather unrelated individuals.  How cooperation can arise from the ``selfishness'' of genes, is one of the most interesting outcome of evolution. 

We have spoken mainly of bacteria, and of asexual reproduction. However, many eukaryotes reproduce sexually. Sex poses a puzzling problem~\cite{Maynard,costofsex}. In true sexual reproduction (not just exchange of genetic material like in some bacteria), the male does offer only genetic information, in form of sperm. The female adds her half of genetic material, plus all the body. The production rate of offspring is therefore proportional to the abundance of females. The sex rate at birth is generally about 50\%, and in comparisons with an asexual replicator or a parthenogenic female, the efficiency is just the half. In spite of this, many organisms that are able to reproduce both sexually and parthenogenycally (many plants, some insects and amphibians) maintain the sexual habits. 

Sex induces also dangerous habits. Especially in large animals, a kind of runoff is observed: one sex develops a phenotypic characteristic, like the tail of peacock, and the other sex finds that the partners with that characteristic are more interesting, and therefore reproduce more. The following generations enhance those characteristics, until the limit (trade-off between survival and exaggeration of the character) is reached.


\subsection{The sequence space}

Let us start our investigation by considering ``standard'' living forms, based on nucleic acid. The storage of information is an one-dimensional structure, possibly fragmented into many chromosomes. We shall call this information the \emph{genome}, and in usual living beings it is formed by one or more filaments of DNA or RNA. Such filaments are formed by a sequence of symbols (basis), and therefore are  digital pieces of information.

Any base can assume one of four values (ATCG in DNA, AUCG in RNA). A sequence of $L$ basis can therefore assume one out of $4^L$ values. The space of all possible sequences is called the sequence space. It is a high-dimensional, discrete space.

Due to the extremely large number of possible sequences, there has not been enough time during evolution to ``try'' all possible sequences nor a relevant fraction of it. Therefore, we cannot apply ``equilibrium'' concepts to evolution. 

The sequence space is well defined for a fixed length $L$, but the length of the genome in real organisms is not fixed. In order to allow for variable $L$, one can consider an extended space with an extra symbol ``*'', which stands for an \emph{empty} space, that can occasionally be replaced by a real basis. This is the same representation used to align sequences, to be illustrated in the following. We can assume that between any subsequent couple of basis there are empty symbols.

\begin{figure}[Ht]
\centerline{\includegraphics[height=0.6\columnwidth,angle=270]{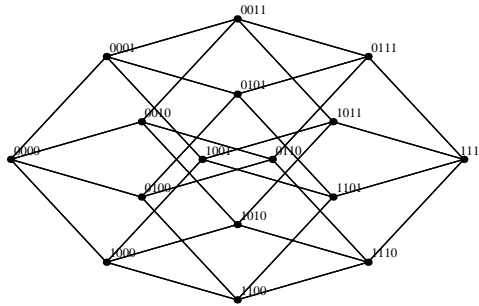}}
\caption{\label{fig:hypercube} The two-dimensional projection of the Boolean hypercube for $L=4$.}
\end{figure}

For simple modeling, we limit our investigation to fixed-length genomes $g = (g_1, \dots, g_L)$, and assume that each component $g_i$ is a Boolean quantity, taking values 0 and 1. The quantity $g_i$ should represent ``genes'' in a rather general meaning. Sometimes, we use $g_i$ to represent two allelic form of a gene, say $g_i=0$ for the wild type, and $g_i=1$ for the less dangerous mutation, even if the quantity is not a ``gene'' in the standard assumption (\emph{e.g.}, information to produce a protein). In other contexts, $g_i$ might represent the presence or the absence of a certain gene, so that we can theoretically identify of all possible genomes with different sequences.

Every possible sequence of length $L$ corresponds to a corner in the Boolean, L-dimensional hypercube (see Figure~\ref{fig:hypercube}). Each link in the hypercube correspond to the change of one element $g_i$, \emph{i.e.}, to a single-element mutation. Other mutations correspond to larger jumps.

In the following, we shall denote by $d(x,y)$ the \emph{genetic distance}, \emph{i.e.},  the minimum number of mutations needed to pass from $x$ to $y$. The quantity $d(x,y)$ is a \emph{topological} concept, in real life one has to consider the probability of mutations in order to estimate if a longer path could occur with higher probability. In our digital approach, $d(x,y)$ is just the Hamming distance between the two sequences. 


\subsection{Mutations}

During lifetime, or in the genome replication phase (when an organism reproduces), some symbol in the genome can be changed: a mutation. Many different types of mutations has been observed. The most common (either because they are actually more common than others, or because they have less chance to cause harmful consequences) are point mutations (an exchange of a basis with another. Other common mutations are deletions (the replacement of a symbol with a ``*'') and insertions (the replacement of a ``*'' with a basis). These mutations correspond to \emph{jumps} is the sequence space. Therefore, a given lineage can be seen as a path in sequence space. By considering all descendants of a given individual (assuming asexual reproduction), we have a set of branching walks, where an ``arm'' stops when an individual dies~\cite{Kauffman87,Kauffman93}.

The probability that a path comes back to an already visited sequence (disregarding the empty symbol) vanishes with growing $L$. Moreover, as we shall see, competition will prevent the merging of different species.

In our model, a mutation is just a change in the sequence $g$, that can occur with a probability that depends only on the sub-sequence that changes, or that may depend on other portion of the genotype. For instance, \emph{point mutations} are the replacement of a zero with an one and vice versa. The probability of this occurrence ($g_i \rightarrow g'_i$) may depend on the position $i$ of the ``gene''. A translocation, on the other hand, involves correlated changes in different positions. Considering the interpretation of $g_i$ as the presence or absence of a certain genetic characteristic, it could be represented as the simultaneous change of two locations $g_i$ and $g_j$ in the genome, that for instance may happen only of $g_i=1$ (presence of the transposable element). Similar representations may be interpreted as gene duplications, etc.

Point mutations correspond to base substitutions. If this happens in a coding region, it causes the change of one codon. Due to the redundancy of genetic code, this mutation may not affect at all the resulting protein, either because the altered codon is read by the same tRNA as the original one, or because it is read by a different tRNA carrying the same amino acid. The modifications in the protein may also be marginal if a amino acid is replaced by another one with similar characteristics.

A deletion or insertion in a coding region causes a ``frame shift'', completely altering the resulting protein. Transpositions also causes similar damages, and the protein is normally not working.\footnote{Some cases of partially-working proteins correspond to genetic diseases, like  Huntington's corea, that allow the survival of affected individuals for some time.} If the protein has a crucial role, this modification lead to the death of the individual. However, it may happen that the same protein is produced by more that one copy of the gene. This generally occurs after a gene duplication, and is vastly more common in eukaryotes (due to the larger genome size and less pressure towards ``efficiency'') than in prokaryotes. In this case, the mutation only reduces the amount of protein produced, a mutation that may be harmful or not. In the latter case, the duplicated gene is ``allowed'' to mutate, generating varied proteins that, inserted in the biochemical network of the cell, may induce new functions (although in general it simply increases the amount of ``junk'' genetic material). This is supposed to be one of the main ``motors'' of evolution, together with the \emph{horizontal} transfer of DNA among individuals (possibly belonging to different species) by retroviruses.\footnote{Transposable elements and gene duplications are often the effects of viral elements, still able to manipulate the genome, but ``trapped'' into the genome by mutations that have destroyed their capability of producing the protective cap.}

In non-coding regions (essentially eukaryotes), the substitutions, insertions and deletions may have ``smoother'' effects, unless they destroy the promoter region of a gene.

In general, we may say that genetic modifications are either lethal, or almost neutral, at least at the level of proteins and biochemical networks. How these almost-neutral modifications (like for instance a change of pigmentation) correspond to the alteration of survival characteristics, depends on their influence on the phenotype, and in how phenotypes of different individuals (and species) interact.

Thus, in many cases one can get rid of the genotypic space: for large scale evolution (formation of species) one can assume that mutations induces a diffusion in the phenotypic space, while for small scale evolution (asymptotic distribution of intra-specific traits) one can assume that mutations are able to generate all possible phenotypic traits.

For simplicity we assume that all point mutations are equally likely, while in reality they depend on the identity of the symbol and on its positions on the genome. For real organisms, the probability of
observing a mutation is quite small. We assume that at most one mutation is possible per generation. We denote with $\mu_s$ the probability of having one point mutation per generation.

The probability to have a point mutation from genotype $y$ to genotype $x$ is given by the short-range mutation matrix $\boldsymbol{M}_s(x,y)$ which is
\begin{equation}
\label{Ms}
  \boldsymbol{M}_s(x|y) =
  \begin{cases}
    1-\mu_s & \text{if $x=y$,} \\
      \dfrac{\mu_s}{L}& \text{if $d(x,y)=1$,} \\
    0&\mbox{otherwise.}
  \end{cases}
\end{equation}
 
Other mutations correspond to long-range jumps in the genotypic space.
A very rough approximation consists in assuming all mutations
are equally probable. Let us denote with $\mu_{\ell}$ the probability per
generation of this kind of mutations. The long-range mutation matrix,
$\boldsymbol{M}_{\ell}$, is defined as
\begin{equation}
\label{Ml}
  \boldsymbol{M}_{\ell}(x|y) =
  \begin{cases}
     1-\mu_{\ell} & \text{if $x=y$,} \\
    \dfrac{\mu_{\ell}}{2^L-1}&\text{otherwise.}
  \end{cases}
\end{equation}

In the real world, only a certain kind of mutations are possible, and in this case  $\boldsymbol{M}_{\ell}$ becomes a sparse matrix $\hat{\boldsymbol{M}}_{\ell}$. We introduce a sparseness index $s$ which is the average number of nonzero off-diagonal elements of $\hat{\boldsymbol{M}}_{\ell}$. The sum of these off-diagonal elements still gives $\mu_{\ell}$. In this case $\hat{\boldsymbol{M}}_{\ell}$ is a quenched sparse matrix, and $\boldsymbol{M}_{\ell}$ can be considered the average of the annealed version.

Both $\boldsymbol{M}_s$ and $\boldsymbol{M}_{\ell}$ are Markov matrices. Moreover, they are circular matrices, since the value of a given element does not depend on its absolute position but only on the distance from the diagonal. This means that their spectrum is real, and that the largest eigenvalue is $\lambda_0=1$. Since the matrices are irreducible, the corresponding  eigenvector $\xi_0$ is non-degenerate, and corresponds to the flat distribution $\xi_0(x) = 1/2^L$.


\subsection{The phenotype}

The phenotype of an individual is how it appears to others, or, better, how it may affect others. Clearly, the phenotype concerns the characteristics of one's body, but may also include other ``extended'' traits, like the dam for beavers, termite nests, or the production of oxygen that finally leaded ancient reductive organisms to extinction.

This function is in general rather complex: genes interact among themselves in an intricate way (epistatic interactions), and while it is easy to ``build'' a bad gene (for instance, a gene that produces a misfolded (incorrectly folded) protein, or a working protein that interferes with the biochemistry of the cell), a ``good'' gene is good only as long there is cooperation with other genes. This is the main reason of the ``general'' failure of genetic engineering: it is difficult to design a gene that produces the desired result (and indeed this is never attempted: one generally tries to move a gene with some effect from an organism to another), but it is yet more difficult to avoid this gene to interfere with the rest of machinery. Only some genetic modifications do not lead to drastic lowering of the output, and most of ``successful stories'' of transgenic plants concerns genes coding for proteins that inhibit some specific poison.

Genes that have additive effects on the phenotype are called ``non-epistatic''; this in general holds only for a given phenotypic trait.

Whithin our modeling, we represent the phenotype of an individual as a set of quantitative characters $u$, that depend on the genotype, the age and on past experiences. A simplification that ease the treatment of the subject consists in assuming that the phenotype is just a function of the genotype:
\[
 u = u(g).
\]
With this assumption, the fitness can be considered a function of the genotype, without introducing the phenotype~\cite{Wright32,Hartle,Peliti95},.

As an example of phenotypes that depend on accumulated characteristics (age), we shall briefly discuss the Penna model in Section~\ref{sec:Penna}.

\subsection{Fitness and the mutation-selection equation}

Selection acts on phenotypes. The fact that an individual survives and reproduces (so perpetuating its genes) may depend on chance. We may however compute (and sometime measure) the ``propensity'' of an individual to survive and produce viable offspring. This quantity is termed ``fitness'', and for our assumptions we may assume that it is proportional to the average number of sons  reaching the reproductive age for a given phenotype $u$, and for a given time interval (generation).

We are now in the position of simulating the evolution of a population. We may represent it as a set of $N$ individuals $x^{(j)}$, $j=1, N$, characterized by their genotype (that unambiguosly characterize the phenotype within our assumptions). The spatial structure of the system may be that of a regular graph (or continuous space with an interaction range), or a social network that may evolve with the system (and in this case may be related to the genotype of the organisms), or a ``well stirred'' environment where all interactions are equally probable.

In each generation, each individual interact with one or more other individuals, accumulating a ``score'' (fitness), that may determine its survival or, equivalently, the probability of producing offspring.

In the simplest arrangement, the time evolution (discrete generations) of an asexual population is given by the following phases:
\begin{enumerate}
  \item Scoring phase: each individual is allowed to interact with others, according with its spatial connectivity (and possibly to displacement like evasion, pursuit, etc.). Their interactions depend on the relative phenotypes, and contribute to the accumulation of a score (generally reset to a default value at the beginning of the phase). The score may be negative or positive.
  \item Survival phase: individuals can survive with a probability that is a monotonic function of the score.
  \item Reproduction phase: empty locations may be colonized by neighboring ones. Colonization implies copying the genome with errors (mutations).
\end{enumerate}

The previous model represents a whole ecosystem, but one is often interested only on some subpart of an ecosystem, like one or few species. If the consistency of a species does not affect others (for instance, one can assume that a species of herbivores does not modify the abundance of grass), then one can neglect to simulate the invariable species, by changing the default value of the score.

Let me be more explicit. Assume that a positive score increases survival, and a negative one decreases it. The presence (interaction) of grass, increases the survival of herbivores, while the encounter with a predator decreases it. If one is interested in the simulation of the interplay among grass, herbivores and  carnivores, then all three species should be simulated. But if one assumes that grass is abundant and not modified by herbivores, one can neglect to include the vegetable substrate, and let herbivore start with a positive score at each step (while carnivores has to meet food to survive)~\cite{Boccara}.

Sexual reproduction implies some modification in the reproductive phase. One can simulate haploid or diploid individuals, with explicit sex determination of just with a recombination among the genomes of the two parents.

Also the displacement may depend on the phenotypic distribution near a given location: preys may try to escape predators, who are pursuing the first. Different sexes may find convenient to stay nearby, and so on.

With the assumption of ``well mixed'' population, \emph{i.e.}, disregarding spatial correlation, and in the limit of very large population the dynamics is described by the mutation-selection  equation~\cite{mutsel} (mean-field approach)
\begin{equation}\label{n}
  n(x, t+1) = \left(1-\frac{N}{K}\right) A(x,\boldsymbol{n}(t))  \sum_{y} M(x|y)  n(y, t+1),
\end{equation}
where $n(x,t)$ is the number of individuals with genome $x$ at time $t$, $M$ is the mutation matrix ($\sum_x M(x|y)=1$) and $A(x,\boldsymbol{n}(t))$ if the fitness of genotype $x$ (or better: of the relative phenotype), given the rest of the population $\boldsymbol{n}(t)$. In the following, we shall neglect to indicate the time dependence, and use a prime to indicate quantities computed one time step further.

The quantity $K$ denotes the carrying capacity, and  $N=\sum_x n(x)$ is
the total number of individuals. The logistic term $1-N/K$ implies that there is no reproduction in the absence of free space, and therefore models the competition among all individual for space. In the limit of populations whose size is artificially kept constant, this term may be included into $A$.

In vector terms, Eq.~\eqref{n} can be written as
\[
 \boldsymbol{n}' = \left(1-\frac{N}{K}\right)\boldsymbol{A}\boldsymbol{M}\boldsymbol{n},
\]
where $\boldsymbol{A}$ is a diagonal matrix.

By summing over $x$, we obtain
\begin{equation}\label{N}
 N' = \ave{A}\left(1-\frac{N}{K}\right)N,
\end{equation}
\emph{i.e.}, the logistic equation, where the reproduction rate $\ave{A}=\sum_x A(x, \boldsymbol{n}) n(x)/N$ depends on the population and therefore on time.

By dividing Eq.~\eqref{n} by Eq.~\eqref{N}, and introducing the frequencies $p(x) = n(x)/N$, we get
\begin{equation}\label{p}
  p'(x) =  \frac{A(x,\boldsymbol{n}(t))}{\ave{A}}\sum_{y} M(x|y)   p(y).
\end{equation}
or
\[
 \boldsymbol{p}' = \frac{\boldsymbol{A}}{\ave{A}}\boldsymbol{M}\boldsymbol{p}.
\]
in the limit of vanishing mutation probability (per generation) and weak selection,  a continuous approach (overlapping generations) is preferred. The mutation-selection equation becomes
\[
 \dot{\boldsymbol{p}} = (\boldsymbol{A}-\ave{A})\boldsymbol{M}\boldsymbol{p}\simeq (\boldsymbol{A}-\ave{A})\boldsymbol{p} + \Delta\boldsymbol{p},
\]
where $\Delta = \boldsymbol{M}-\boldsymbol{1}$, and $\boldsymbol{1}$ is the identity. Considering only symmetric point mutations, $\Delta$ is the $L$-dimensional diffusion operator. One can recognize here the structure of a reaction-diffusion equation: evolution can be considered as a reaction-diffusion process in sequence space. The typical patterns are called \emph{species}.

This simple model allows to recover some known results.

\section{Evolution on a fitness landscape}

The analysis of the replication equation, Eq.~\eqref{p} is much simpler if the fitness $A$ does not depend on the population structure, \emph{i.e.}, $A=A(x)$. The lineage of an individual corresponds to a walk on a static landscape, called the \emph{fitness landscape}. A generic fitness function may be considered a landscape for short times, in which the species, other than the ones under investigations, can be considered constant.

In this case the evolution really corresponds to an optimization process: in the case of infinite population,  and for mutations that are able to ``connect'' any two genomes (in an arbitrary number of passes -- $M$ is an irreducible matrix), the system may reach an equilibrium state (in the limit of infinite time). The mutation-selection equation may be linearized by considering that the relative probability $q(\boldsymbol{x})$ of a path $\boldsymbol{x}$ (composed by a time sequence of genotypes $\boldsymbol{x} = x^{(1)}, x^{(2)}, \dots, x^{(t)}$) is
\begin{equation}\label{q}
 q(\boldsymbol{x}) = \frac{
  \prod_{t=1}^{T+1} M(g^{(t+1)}|x^{(t)}) A(x^{(t)})
 }{
  \prod_{t=1}^{T+1} \ave{A}(t)
 },
\end{equation}
where the denominator is common to all paths. Thus, the unnormalized probabilities of genotypes at time $t$,  $z(x)\equiv z(x,t)$, evolve as
\begin{equation}\label{z}
 z'(x) = \sum_{y} M(x|y) A(y) z(y) \qquad \text{or}\qquad  \boldsymbol{z}' = \boldsymbol{M}\boldsymbol{A}\boldsymbol{z}.
\end{equation}
Notice that $z(x,t)$ can be considered as a restricted partition function, see Section~\ref{sec:StatMech}. 

The structure of probabilities in Eq.~\eqref{q} suggests to use an exponential form for $A$:
\[
 A(x, \boldsymbol{p}) = \exp(H(x, \boldsymbol{p})),
\]
that makes evident the analogy with statistical mechanics: the fitness landscape behaves as a sort of energy, while mutations act like temperature~\cite{Leuthausser,Tarazona,Galluccio}. 
See also Section~\ref{sec:StatMech}.

There are a few landscapes that have been studied in details\cite{Peliti:Introduction}. The simplest one is the flat landscape, where selection plays no role. It is connected to the concept of \emph{neutral evolution}, that we shall examine in Section~\ref{sec:Neutral}. In this landscape, evolution is just a random walk in sequence space, and one is interested in the probability of fixation of mutations in a finite populations, which corresponds to the divergence of an isolated bunch of individuals (allopatric speciation).

A species is in general a rather stable entity, in spite of mutations. One can assume that it occupies a \emph{niche}, which corresponds to a maximum of the fitness in the phenotypic space. The presence of this niche generally depends on other species, like for instance insects specialized in feeding on a particular flower have their niche tied to the presence of their preferred food. An extremal case is that of a \emph{sharp} fitness landscape constituted by an isolated peak. This landscape is particular since it is not possible to guess where the top of the fitness is, unless one is exactly on it.
A lineage in a flat genotypic space is similar to a random walk, due to mutations. The only possibility to find the top is though casual encounter, but is a high-dimensional space, this is an extremely unfavorable event. 

For small mutation probability, the asymptotic distribution is a \emph{quasispecies} grouped around the peak. By increasing the mutation probability (or equivalently the genome length), this cloud spreads. It may happen that in finite populations no one has the right phenotype, so that the peak is lost.  This is the \emph{error threshold} transition, studied in Section~\ref{sec:ErrorThreshold}. Clearly, this transition poses the problem of how this peak has been populated for the first time. The idea (see Section~\ref{SOC}), is that the absence of population fitness space is actually similar to a Swiss cheese: paths of flat fitness and ``holes'' of unviable phenotypes (corresponding essentially to proteins that are unable to fold). The ``corrugation'' of the fitness is due to the presence of other species. So for instance a highly specialized predator may become so tied to its specific prey, that its effective fitness landscape (for constant prey population) is extremely sharp.  

Another consequences of an increased mutation rate, more effective for individuals with accurate replication machinery like multicellular ones, is the extinction of the species without losing their ``shape'', the so-called  Muller's ratchet~\cite{Lynch90,Lynch93,Bernardes} or stochastic escape~\cite{Higgs95,Wood96}, which, for finite populations, causes the loosing of fitter strains by stochastic fluctuations. Muller's ratchet is studied in Section~\ref{sec:ErrorThreshold}.

Genes that have an additive effect (non-epistatic) lead to \emph{quantitative traits}, and to smoother landscape, shaped like the Fujiyama mount. It is possible to obtain a good approximation for the asymptotic distribution near such a maximum, as shown in Section~\ref{sec:SmoothLandscape}. Such a results will come at hand in dealing with competition, Section~\ref{sec:PhenotypicSpeciation}. On a Fujiyama landscape, no error threshold transition is present~\cite{Peliti:Introduction}.

Finally, one can study the problem of the evolution in a landscape with variable degree of roughness, a problem similar to that of disordered media in statistical mechanics~\cite{Peliti,Peliti95,Peliti:Review}. 

A rather generic formulation follows lines similar to that of disordered statistical problems~\cite{Peliti:Introduction}.  The phenotype $u(x)$ of a genotype $x=(x_1, x_2, \dots, x_L)$ is given by 
\begin{equation}
  u(x) = \sum_{\{i_1\dots i_K\}} J_{\{i_1\dots i_K\}} x_{i_1}\cdots x_{i_K},
  \label{pspin}
\end{equation}
where     $J_{\{i_1\dots i_K\}}$, for each different set of indexes $\{i_1\dots i_K\}$, are independent, identically distributed, random variables, so that for every $K$ we are dealing with a random ensemble of phenotypes (and of fitness). The larger $K$, the faster the fitness correlations decay in sequence space, so that the fitness landscape is less and less correlated, \emph{i.e.}, as it is usually said, more and more rugged. In the limit $k\rightarrow\infty$, one has a complete disordered fitness landscape, where the values of the fitness at different positions in sequence space are independent random variables  (random energy model~\cite{Derrida}). For $K=2$ this model is similar to the one introduced by Kauffman~\cite{Kauffman} for modeling the genetic network of a cells. 

On an extremely rugged landscape one observes an error threshold transition~\cite{Peliti:REM,Peliti:Introduction} between a state in which the population distribution is concentrated on local maxima (localized phase), and a wide distribution. Due to the disorder, the distribution in the localized phase depends on the initial conditions, \emph{i.e.}, on the past history of the population.


\subsection{Evolution and optimization, replicator equation}
The evolution may appear as an optimization process. Indeed, if we neglect mutations, Eq.~\eqref{p}
becomes
\[
 \boldsymbol{p}' = \frac{\boldsymbol{A}}{\ave{A}}\boldsymbol{p},
\]
called the \emph{replicator equation} (discrete time). 

Assume that we start from a uniform distribution over the whole phenotypic space, and that the fitness shows a single, smooth maximum. The phenotypes $u$ with $A(u)<\ave{A}$ tend to decrease in frequency, while those with $A(u)>\ave{A}$ tend to increase their frequency. Because of this, the average fitness $\ave{A}$ increases with time (Fisher theorem~\cite{Fisher}). Notice that this results may be compatible with the extinction of the population, either due to chaotic oscillations for high reproductive rates, oscillations that may substantially reduce the population to a level in which random sampling is important, or because $\ave{A}$ is below one in Eq.~\ref{N}. An example of this is given by genes that alter the sex ratio. These genes are able to ``kill'' male embryos, and thus increase their own population (since they reside on the female chromosome) although the final fate of the populatiion is extinction. This effect may be the origin of the smallness of male Y chromosome in mammals: a smaller chromosome implies less genes that may be targeted by the proteins produced by the killer genes.

The structure of Eq.~\eqref{p} says that the fitness does not have absolute values, except that the average fitness (the basic reproductive rate) has to be greater than one. For a given genome, the survival of individuals depends on their \emph{relative} fitness: those that happen to have a fitness larger than average tend to survive and reproduce, the others tend to disappear. By doing so, the average fitness in general increases, so that a genome that is good at a certain time will become more common and the relative fitness more similar to the average one. The effects of this generic tendency depend on the form of the fitness. For instance,  for offspring production, it is more convenient to invest in females than in males. But as soon as males become rare, those that carry a gene that increases their frequency in the population will be more successful, in that the females fertilized with this gene will produce more males, that in average will fertilize more females and so on.

The presence of sexual reproduction itself is difficult to be interpreted as an optimization process. As reported in the Introduction, the offspring production of sexual species is about a half of that of asexual ones. The main advantage of sexual reproduction (with diploidicity) is that of maintaining a genetic (and therefore genotypic) diversity by shuffling paternal and maternal genes. This has the effect of maintaining genetic and phenotypic diversity, without the risk associated with a high mutation rate: the generation of unviable offspring, that lead to the error threshold or to the mutation meltdown, Section~\ref{sec:ErrorThreshold}. This genetic diversity is useful in colonizing new environments or for variable ones~\cite{BagnoliGuardiani}, but is essential to escape the exploitation from parasites, that reproduce (and therefore evolve) faster than large animals and plants~\cite{Hamilton}.

Therefore, sex can be considered an optimization strategy only for variable environments. The fact that an optimization technique inspired by sexual reproduction, genetic algorithms~\cite{Holland}, is so widely used is somewhat surprising.

\subsection{Evolution and statistical mechanics}\label{sec:StatMech}

When one takes into consideration only point mutations ($\boldsymbol{M}\equiv \boldsymbol{M}_s$),
Eq.~(\ref{z}) can be read as 
the transfer matrix of a two-dimensional Ising
model~\cite{Leuthausser,Tarazona,Baake}, 
for which the genotypic element $x_i^{(t)}$ corresponds to the spin in
row $t$ and column $i$, and $z(x,t)$ is the restricted partition function of
row $t$.  The effective Hamiltonian (up to  constant terms) 
of a possible lineage  $\boldsymbol{x}=(x^{(t)})_t=1^T$ from time $1\le t \le T$ is
\begin{equation}
  \mathcal{H} = \sum_{t=1}^{T-1} \left(\gamma \sum_{i=1}^L
  x_i^{(t)} x_{i}^{(t+1)} + H\bigl(x^{(t)}\bigr)\right),
  \label{ising} 
\end{equation}
where $\gamma=-\ln(\mu_s/(1-\mu_s))$.
 
This peculiar two-dimensional Ising model has a long-range coupling along
the row (depending on the choice of the fitness function) and a
ferromagnetic coupling along the time direction (for small short
range mutation  probability).  In order to obtain the statistical
properties of the system one has to sum over all possible
configurations (stories), eventually selecting the right boundary
conditions at time $t=1$. 

The bulk properties of Eq.~(\ref{ising})  cannot be reduced in general
to the equilibrium distribution of an one-dimensional system, since the transition probabilities among rows do
not obey detailed balance. Moreover, the temperature-dependent
Hamiltonian (\ref{ising}) does not allow an easy identification
between energy and selection, and temperature and mutation, what is
naively expected by the biological analogy with an adaptive walk. 

A Ising configuration of Eq.~(\ref{ising}) corresponds to a possible
genealogical story, \emph{i.e.}, as a directed polymer in the genotypic
space~\cite{Galluccio}, where mutations play the role of elasticity.
It is natural to try to rewrite the model in terms of the sum over all
possible paths in genotypic space. 

As shown in Ref.~\cite{Bagnoli:SmallWorld}, in the case of long-range mutations, this scenario simplifies, and the asymptotic probability distribution  
$\tilde{\boldsymbol{p}}$
is  proportional to the diagonal of $\boldsymbol{A}^{1/\mu_{\ell}}$:
\begin{equation}
  \tilde p(x) = C \exp\left(\dfrac{H(x)}{\mu_{\ell}}\right),
  \label{Boltzmann}
\end{equation}
\emph{i.e.}, a Boltzmann distribution with Hamiltonian $H(x)$ 
and temperature $\mu_{\ell}$. This corresponds to the naive analogy between
evolution and equilibrium statistical mechanics. In other words, the
genotypic distribution is equally populated if the phenotype is the
same, regardless of the genetic distance since we used long-range
mutations. The convergence to equilibrium is more rapid for rough landscapes. 

For pure short-range mutations, this correspondence holds only approximately for very weak selection or very smooth landscapes. The reason is that in this case the coupling in the time direction are strong, and mutations couple genetically-related strains, that may differ phenotypically. 

\begin{figure}
  \centerline{\includegraphics[width=0.8\columnwidth]{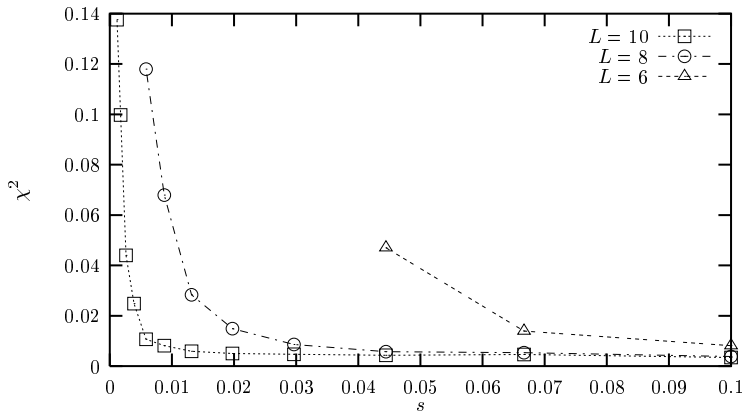}}
  \caption{Scaling of $\chi^2$ with the sparseness factor $s$, for
  three values of genome length $L$. $\mu_{\ell}=0.1$ and  $\mu_s=0.1$.} 
  \label{fig:sparsechi} 
\end{figure}

Since short and long-range mutations have an opposite effect, it is interesting to study the more realistic case in which one has more probable short-range mutations, with some long-range ones that occur sporadically, always between the same genomes (quenched disorder). This scenario is very reminiscent of the small-world phenomenon~\cite{Watts:SmallWorld}, in which a small percentage $s$ (sparseness) of long-range links added to a locally connected lattice is able to change its diffusional properties. Even in the limit of $s\rightarrow 0$ (after long times), one observes a mean-field (well-stirred) distribution. We computed the variance $\chi^2$ of the deviations of the asymptotic distribution and that of Eq.~\eqref{Boltzmann} for a rough landscape. As can be seen in Figure~\ref{fig:sparsechi}, for large genomes the transition to the mean-field limit occurs for $s\rightarrow 0$.

\subsection{Quasispecies, error threshold and Muller's ratchet}\label{sec:ErrorThreshold}

Before going in deep studying a general model of an evolving ecosystem
that includes the effect of competition (co-evolution), let us discuss
a simple model~\cite{Bagnoli:ijmpc} that presents two possible mechanisms of
escaping from a local optimum, \emph{i.e.}  the error threshold and the Muller's ratchet.

We consider a \emph{sharp peak landscape}: the 
phenotype $u_0=0$, corresponding to the master sequence 
genotype $x=0\equiv(0,0,\dots)$ 
has higher fitness $A_0=A(0)$, and all 
other genotypes have the same, lower, fitness $A_*$.  
Due to the form of the fitness function, the dynamics of the
population is fundamentally determined by the fittest strains.

Let us indicate
with $n_0=n(0)$ the number of individuals sharing the master sequence, 
 with $n_1=n(1)$ the
number of individuals with phenotype $u=1$ (only one bad gene, i.e. 
a binary string with all zero, except a single $1$), and with
$n_*$ all
other individuals. 
We assume also non-overlapping generations, 

During reproduction, individuals with phenotype $u_0$ can mutate, 
contributing to $n_1$, and those with phenotype $u_1$ can mutate, 
increasing  $n_*$. 
We disregard the possibility of back mutations from
$u_*$ to $u_1$ and from $u_1$ to $u_0$. 
This last assumption is
equivalent to the limit $L\rightarrow \infty$, which is the case for existing
organisms. We consider only short-range mutation with probability
$\mu_s$.
Due to the assumption of large $L$, the multiplicity factor of
mutations from $u_1$ to $u_*$ (i.e. $L-1$)
is almost the same of that from $u_0 $
to $u_1$ (i.e. $L$).  
 
The evolution equation Eq~\eqref{n} of the population becomes
\begin{equation}
\begin{split}
 n'_0 &=\left(1-\dfrac{N}{K}\right)(1-\mu_s) A_0 n_0, \\
 n'_1 &=\left(1-\dfrac{N}{K}\right)\bigl(
  (1-\mu_s) A_* n_1+\mu_s A_0 n_0\bigr),\\
 n'_* &=\left(1-\dfrac{N}{K}\right)  A_* (n_*+\mu_s n_1).
\end{split}
\label{toy:n}
\end{equation}
and 
\[
  \ave{A}=\dfrac{A_0 n_0 + A_* (n_1+n_*)}{N}
\]
is the average fitness of the population.

The steady state of Eq.~(\ref{toy:n}) 
is given by $\boldsymbol{n}'=\boldsymbol{n}$. 
There are three possible fixed points $\boldsymbol{n}^{(i)}=\left(n_0^{(i)},n_1^{(i)},n_*^{(i)}\right)$:
 $\boldsymbol{n}^{(1)} = (0,0,0)$ ($N^{(1)}=0$), $\boldsymbol{n}_2 = (0,0,K(1-1/\ave{A_*}))$ ($N^{(2)}=n_*^{(2)}$) and
\[
  \boldsymbol{n}^{(3)} = \begin{cases}
    n_0^{(3)} &=N^{(3)}\dfrac{(1-\mu_s) A_0 -A_*}{A_0 -A_*}, \\
    n_1^{(3)} &=N^{(3)}\dfrac{\mu_s}{1-\mu_s}\dfrac{ A_0(q A_0
      -A_*)}{(A_0 -A_*)^2}, \\
    n_*^{(3)} &=N^{(3)}\dfrac{\mu_s^2}{1-\mu_s} \dfrac{A_0A_*}{(A_0
      -A_*)^2},\\
    N^{(3)} &= 1-\dfrac{1}{A_0(1-\mu_s)}.
  \end{cases}
\]

The fixed point $\boldsymbol{n}^{(1)}$ corresponds to extinction of the whole population, i.e.\ to mutational meltdown (MM). It is trivially stable if $A_0<1$, but it can become stable also if $A_0>1$, $A_* <1$  and
\begin{equation}
  \mu_s > 1-\dfrac{1}{A_0}.
  \label{MR}
\end{equation} 

The fixed point $\boldsymbol{n}^{(2)}$ corresponds to a distribution in which the master sequence has disappeared even if it has larger fitness than other phenotypes. This effect is usually called Muller's ratchet (MR). The point $P_2$ is stable for $A_0>1$,  $A_* >1$ and
\begin{equation}
  \mu_s > \dfrac{A_0/A_*-1}{A_0/A_*}.
  \label{MM}
\end{equation}

The fixed point $\boldsymbol{n}^{(3)}$ corresponds to a coexistence of all phenotypes. It is stable in the rest of cases, with $A_0>1$. The asymptotic distribution, however, can assume two very different shapes. In the quasi-species (QS) distribution, the master sequence phenotype is more abundant than other phenotypes; after increasing the mutation rate, however, the numeric predominance of the master sequence is lost, an effect that can be denoted error threshold (ET). The transition between these two regimes is given by $n_0=n_1$, \emph{i.e.},
\begin{equation}
  \mu_s = \dfrac{A_0/A_*-1}{2A_0/A_*-1}. 
  \label{ET}
\end{equation}
Our definition of the error threshold transition needs some remarks: in Eigen's original work~\cite{Eigen71,Eigen:quasispecies} the error threshold is located at the maximum mean Hamming distance, which corresponds to the maximum spread of population. In the limit of very large genomes these two definitions agree, since the transition becomes very sharp~\cite{Galluccio}. See also Refs.~\cite{Baake1,Baake3}.

\begin{figure}[Ht]
\centerline{
\includegraphics[width=0.8\columnwidth]{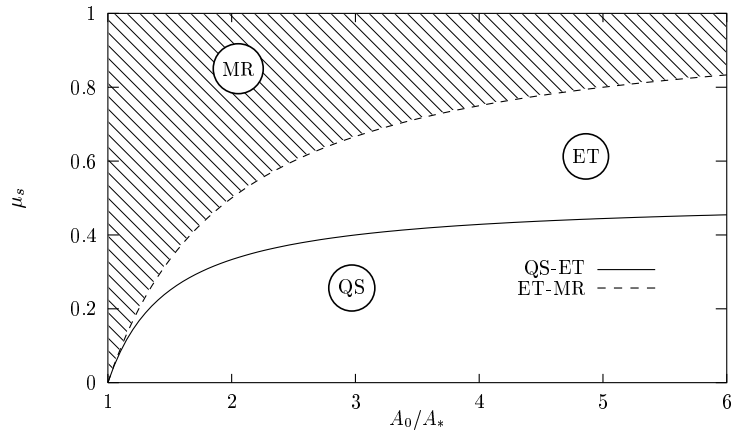}
}
\caption{Phase diagram for the error threshold and Muller's ratchet
transitions ($A_*>1$).
 MR refers to the Muller's ratchet phase
ET  to the error threshold distribution and QS to
 quasi-species distribution. The phase boundary between the 
 Muller's ratchet effect and
 the error threshold distribution (Eq.~(\protect{\ref{MR}})) is marked ET-MR;
the phase boundary between 
 the error threshold and the quasi-species 
 distribution (Eq.~(\protect{\ref{ET}})) is marked QS-ET. 
 }
\label{fig:toy1}
\end{figure}

In Figure~\ref{fig:toy1} we reported the phase diagram of model (\ref{toy:n}) for $A_*>1$ (the population always survives). There are three regions:  for a low mutation probability $\mu_s$ and high selective advantage $A_0/A_*$ of the master sequence, the distribution has the quasi-species form (QS); increasing $\mu_s$ the distribution undergoes the error threshold (ET) effect; finally, for very high mutation probabilities, the master sequence disappears and we enter the Muller's ratchet (MR) region~\cite{Malarz,Baake}.
The error threshold phase transition is not present for smooth landscapes (for an example of a study of evolution on a smooth landscape, see
Ref.~\cite{Kessler}).

\begin{figure}[Ht]
\centerline{
\includegraphics[width=0.8\columnwidth]{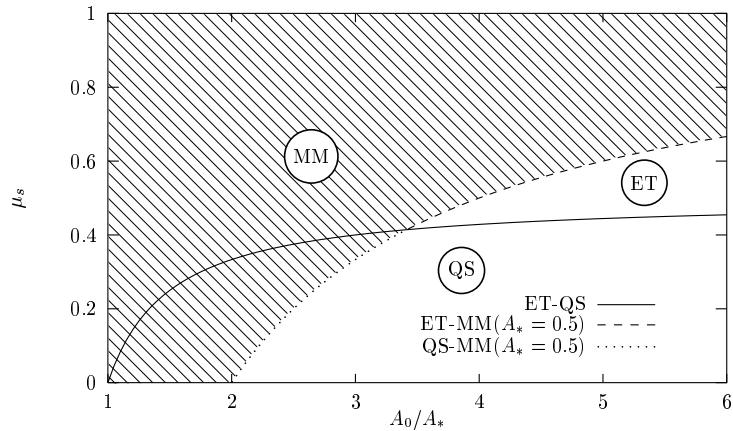}
}
\caption{Phase diagram for the mutational meltdown extinction,
 the error threshold and the
quasi-species distributions ($A_*<1$).
 MM refers to the mutational meltdown phase,
ET  to the error threshold distribution and QS to
 quasi-species distribution. The phase boundary between the 
 Mutational meltdown effect and
 the error threshold distribution (Eqs.~(\protect{\ref{MM}})
 and (\protect{\ref{ET-MM}})) 
 is marked ET-MM;
the phase boundary between 
 the mutational meltdown and the quasi-species 
 distribution (Eqs.~(\protect{\ref{MM}}) and (\protect{\ref{QS-MM}})) 
 is marked QS-MM. 
 }
\label{fig:toy2}
\end{figure}

In Figure~\ref{fig:toy2} we illustrate the phase diagram in the case $A_*=0.5$. For a low mutation probability $\mu_s$ and high selective advantage $A_0/A_*$ of the master sequence, again one observes a quasi-species distribution (QS), while for sufficiently large $\mu_s$ there is the extinction of the whole population due to the mutational meltdown (MM) effect.  The transition between the QS and MM phases can occur directly, for 
\begin{equation}
  A_0/A_*< \dfrac{1-\sqrt{1-A_*}}{A_*}
  \label{QS-MM}
\end{equation}
 (dotted QS-MM line in Figure): 
during the transient before
extinction the distribution keeps the QS form. For  
\begin{equation}
  A_0/A_* > \dfrac{1-\sqrt{1-A_*}}{A_*}
  \label{ET-MM}
\end{equation}
one has first the error threshold transition (QS-ET line in Figure), and then one observes extinction due to the mutational meltdown effect (dashed ET-MM line in Figure). This mutation-induced extinction has been investigated numerically in Ref.~\cite{Malarz}. 

\begin{figure}[Ht]
\centerline{
\includegraphics[width=0.8\columnwidth]{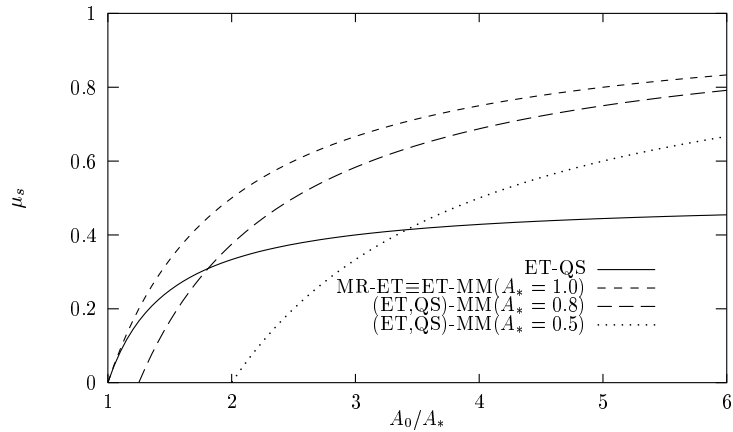}
}
\caption{Phase diagram for the error threshold and mutational meltdown
transitions for some values of $A_*$. 
ET-QS refers to the Error threshold transition,
 Eq.~(\protect\ref{ET}),
QS-MM to the mutational meltdown extinction without the error threshold
transition, 
 Eqs.~(\protect\ref{MM}) and (\protect{\ref{QS-MM}}),
ET-MM to the mutational meltdown extinction after the error threshold
transition, Eqs.~(\protect\ref{MM}) and  (\protect{\ref{ET-MM}}).
The line MR-ET marks the Muller's ratchet  boundary,  Eq.~(\protect\ref{MR}),
which  coincides with   
the mutational meltdown (MM) boundary for $A_*=1$. 
 }
\label{fig:toy3}
\end{figure}

We finally report in Figure~\ref{fig:toy3} the phase diagram of the model in the
$A_*<1$ case, for some values of $A_*$. Notice that for $A_*=1$ the mutational
meltdown effect coincides with the Muller's ratchet one. 

Here the mutation probability $\mu_s$ is defined on a per-genome basis. If one
considers a fixed mutation probability $\mu$ per genome element, one 
has $\mu_s \simeq L \mu$, where $L$ is the genome length. 
Thus, it is possible to trigger 
these phase transitions by increasing the genome length.  

Numerical simulations on a related model~\cite{Malarz} does not show a phase in which, for $A_*<1$, the population survives after the loss of the quasispecies distribution, \emph{i.e.}, the error theshold - mutational meltdown (ET-MM) transition was not observed. 

The Error threshold
for finite populations has been studied in
Refs.~\cite{NowakSchuster,Baake,Alves}.

\subsection{Evolution in a phenotypic space}
\label{sec:PhenotypicEvolution}

In the previous section we have obtained some information about the
stability of a quasi-species distribution. In the following we want to
study the stability of a distribution formed by more than one
quasi-species, i.e.\ the speciation phenomenon. Before doing that we
need to know the shape of a quasi-species given a static fitness
landscape. Some analytical results can be obtained by 
considering the dynamics only in the phenotypic space~\cite{Kessler}. 

We assume that the phenotypic index $u$ ranges between $-\infty$ and
$\infty$
in unit steps (the fitness landscape provides that only a finite range
of the phenotypic space is viable), and  that  
mutations connect 
phenotypes at unit distance; 
the probability of observing a mutation per unit of time is $\mu$. 
The mutational matrix $M(u,v)$ has the form:
\[
   M(u,v) = 
   \left\{\begin{array}{ll}
    \mu & \mbox{if $|u,v|=1$,} \\
    1-2\mu & \mbox{if $u=v$,}\\
    0&\mbox{otherwise.}
  \end{array}\right.
\]

Let us  consider as before the evolution of phenotypic distribution $p(u)$,
that gives the probability of observing the phenotype $u$. 
As before the whole distribution is denoted by $\boldsymbol{p}$.


Considering a phenotypic linear space and non-overlapping generations, we get from Eq.(\ref{p})

\[
  p'(u)=
  \dfrac{(1- 2 \mu)A(u,\boldsymbol{p})p(u) + \mu (A(u+1,\boldsymbol{p})p(u+1)+
  A(u-1,\boldsymbol{p})p(u-1)}{\ave A}.   
\]

In the limit of continuous phenotypic space, $u$ becomes a real number and 
\begin{equation}
p'(u)= \dfrac{1}{\ave{A}}\left(
  A\bigl(u,\boldsymbol{p}\bigl)p(u) 
  +\mu \dfrac{\partial^2A\bigl(u,\boldsymbol{p}\bigl)p(u)}{\partial
  u^2}
 \right), \label{evol1}
\end{equation}

with
\begin{equation}
\int_{-\infty}^{\infty} p(u)du =1, \qquad \int_{-\infty}^{\infty} A(u,\boldsymbol{p})p(u) du=\ave{A}.
\label{norm}
\end{equation}

Eq.~(\ref{evol1}) has the typical form of 
 a nonlinear reaction-diffusion equation.
The numerical solution of this equation shows that a stable asymptotic 
distribution exists for almost all initial conditions.

The fitness $A(u,\boldsymbol{p})=\exp (H(u,\boldsymbol{p}))$ can be written 
as before, with
\[
  H(u,\boldsymbol{p})=V(u)+\int_{-\infty}^{\infty} J(u,v) p(v) dv .\label{fitness}
\]

Before studying the effect of competition and the speciation transition
let us derive the exact form of
$p(u)$ in case of a smooth and sharp static fitness landscape.

\subsubsection{Evolution near a smooth and sharp maximum} 
\label{sec:SmoothLandscape}

In the presence of a single maximum  
the asymptotic distribution is given by one quasi-species
centers around the global maximum of the static landscape. 
The effect of a finite mutation rate is simply that of broadening
the distribution from a delta peak to a bell-shaped curve.

We are interested in deriving the exact form of the 
asymptotic distribution near the maximum. 
We take a static fitness $A(u)$
with a smooth, isolated maximum for $u=0$ ({\it smooth maximum} approximation).
Let us assume that 
\begin{equation}
  A(u)\simeq A_{0}(1-au^{2}), 
  \label{pot}
\end{equation}
where $A_0 = A(0)$.
Substituting $\exp (w)=Ap$ in Eq.~(\ref{evol1}) we have (neglecting to indicate
the phenotype  $u$, and using primes to denote differentiation with respect
to it): 
\[
  \dfrac{\ave{A} }{A}=1+\mu ({w'}^{2}+w''), 
\]
 and approximating $A^{-1}=A_0^{-1}\left( 1+au^{2}\right) $, we have 
\begin{equation}
\dfrac{\ave{A} }{A_{0}}(1+au^{2})=1+\mu ({w'}^{2}+w''). \label{alphaA0}
\end{equation}
A possible solution is 
\[
  w(u)=-\dfrac{u^{2}}{2\sigma ^{2}}. 
\]
Substituting into Eq. (\ref{alphaA0}) we finally get 
\begin{equation}
  \dfrac{\ave{A} }{A_{0}}=\dfrac{2+a\mu - 
  \sqrt{4a\mu +a^{2}\mu ^{2}}}{2}.    
  \label{smooth:app}
\end{equation}
Since $\ave{A} /A_{0}$ is less than one we have
 chosen the minus sign. In the limit $a\mu \rightarrow 0$ (small mutation rate
and smooth maximum), we have 
\begin{equation}
  \dfrac{\ave{A} }{A_{0}}\simeq 1-\sqrt{a\mu } \label{aveA}
\end{equation}
and
\begin{equation}
  \sigma ^{2}\simeq \sqrt{\dfrac{\mu }{a}}. \label{sigma}
\end{equation}

The asymptotic solution is 
\begin{equation}
  p(u)= \dfrac{1+au^{2}}{\sqrt{2\pi }\sigma (1+a\sigma ^{2})}\exp \left(
  -\dfrac{u^{2}}{2\sigma ^{2}}\right), \label{phenotypicp}
\end{equation}
so that $\int p(u)du=1$. The solution is then a bell-shaped curve, its
width $\sigma$ being determined by the combined effects
of the curvature $a$ of maximum and the mutation rate $\mu$.

For completeness, we study here also the case of a {\it sharp maximum},  
for which  $A(u)$ varies considerably with $u$. In this case 
the growth rate of less fit strains has a 
large contribution from the mutations of fittest strains, 
while the reverse flow is negligible, thus
\[
  p(u-1)A(u-1) \gg p(u)A(u) \gg p(u+1)A(u+1) 
\]
neglecting last term, and substituting  $q(u)=A(u)p(u)$ in Eq.~(\ref{evol1})
we get: 
\begin{equation}
  \dfrac{\ave{A}}{A_0}  = 1-2\mu \qquad \mbox{for $u=0$}\label{map0}
\end{equation}
and 
\begin{equation}
  q(u) =\dfrac{\mu}{\left(\ave{A} A(u) 
  -1+2\mu\right)} q(u-1) \qquad \mbox{for
  $u>0$} \label{map}
\end{equation}

Near $u=0$, combining Eq.~(\ref{map0}), Eq.~(\ref{map})and Eq.~(\ref{pot})), 
we have
\[
  q(u) =\dfrac{\mu}{(1-2\mu)a u^{2}} q(u-1). 
\]

In this approximation the solution is
\[
  q(u) = \left(\dfrac{\mu}{1-2\mu a}\right)^u \dfrac{1}{(u!)^2},
\]
and 
\[
  y(u) = A(u)q(u) \simeq \dfrac{1}{A_0}(1+a u^2)
  \left(\dfrac{\mu A_0}{\ave{A}
  a}\right)^u \dfrac{1}{u!^2}. 
\]

\begin{figure}[Ht]
\includegraphics[width=0.8\columnwidth]{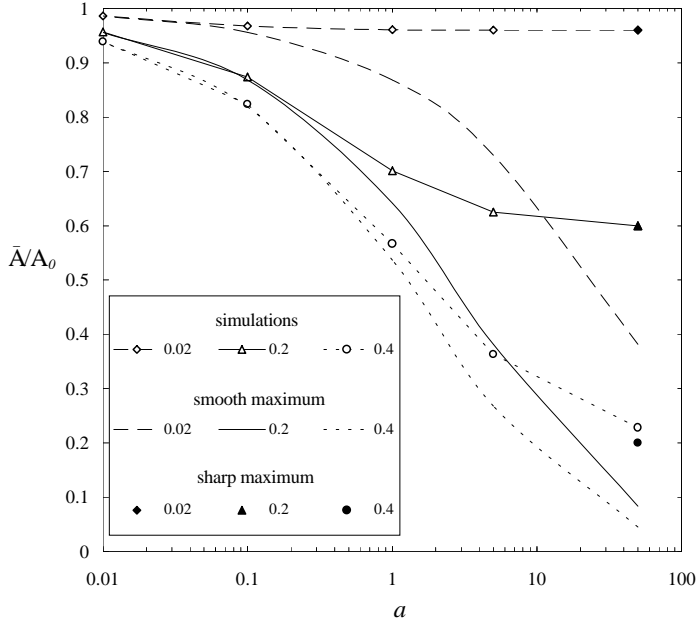}
\caption{Average fitness $\ave{A}/A_{0}$ versus the coefficient $a$,
of the fitness function, Eq.~(\ref{pot}),  for some values of the mutation
rate $\mu$. Legend: {\it numerical solution} corresponds to the numerical solution of 
Eq.~(\ref{evol1}),  
{\it smooth maximum} refers to  Eq.~(\ref{smooth:app}) and {\it sharp maximum}
to Eq.~(\ref{map0})}
\label{alpha}
\end{figure}

We have checked the validity of these approximations by solving numerically   
Eq.~(\ref{evol1}); 
the comparisons are shown 
in Figure~(\ref{alpha}).
We observe that the {\it smooth maximum} approximation agrees with the 
numerics for small values of $a$, 
 when $A(u)$ varies slowly with $u$, while the {\it
sharp maximum} approximation  agrees with the  numerical results for large
 values of  $a$, when small variations of $u$ correspond to large
variations of $A(u)$.


\subsection{Coexistence on a static fitness landscape}
\label{sec:Coexistence}

We investigate here the conditions for which more than one
quasi-species can coexist  on a static
fitness landscape without competition.

Let us assume that the fitness landscape has several distinct peaks,
and that any peak can be approximated by a quadratic function near its
maximum. 
For
small but finite mutation rates, as shown by Eq.~(\ref{phenotypicp}),
the distribution around an isolated maxima is a bell curve, whose
width is given by Eq.~(\ref{sigma}) and average fitness by
Eq.~(\ref{aveA}). Let us call thus distribution a quasi-species, and
the peak a niche.

If the niches are separated by a distance greater
than $\sigma$, a superposition of quasi-species~(\ref{phenotypicp}) 
is a solution of 
Eq.~(\ref{p}). Let us number the quasi-species with the
index $k$:
\[
  p(u) = \sum_k p_k(u);
\]
each $p_k(u)$ is centered around $u_k$ and has average fitness
$\ave{A}_k$.  The condition for the coexistence of two
quasi-species $h$
and $k$ is
$\ave{A}_h = \ave{A}_k$ (this condition can be extended to
any number of quasi-species). 
In other terms one can
say that in a stable environment the fitness of all co-existing individuals is the same,
independently on the species.

\begin{figure}[Ht]
\centerline{\includegraphics[width=0.35\columnwidth]{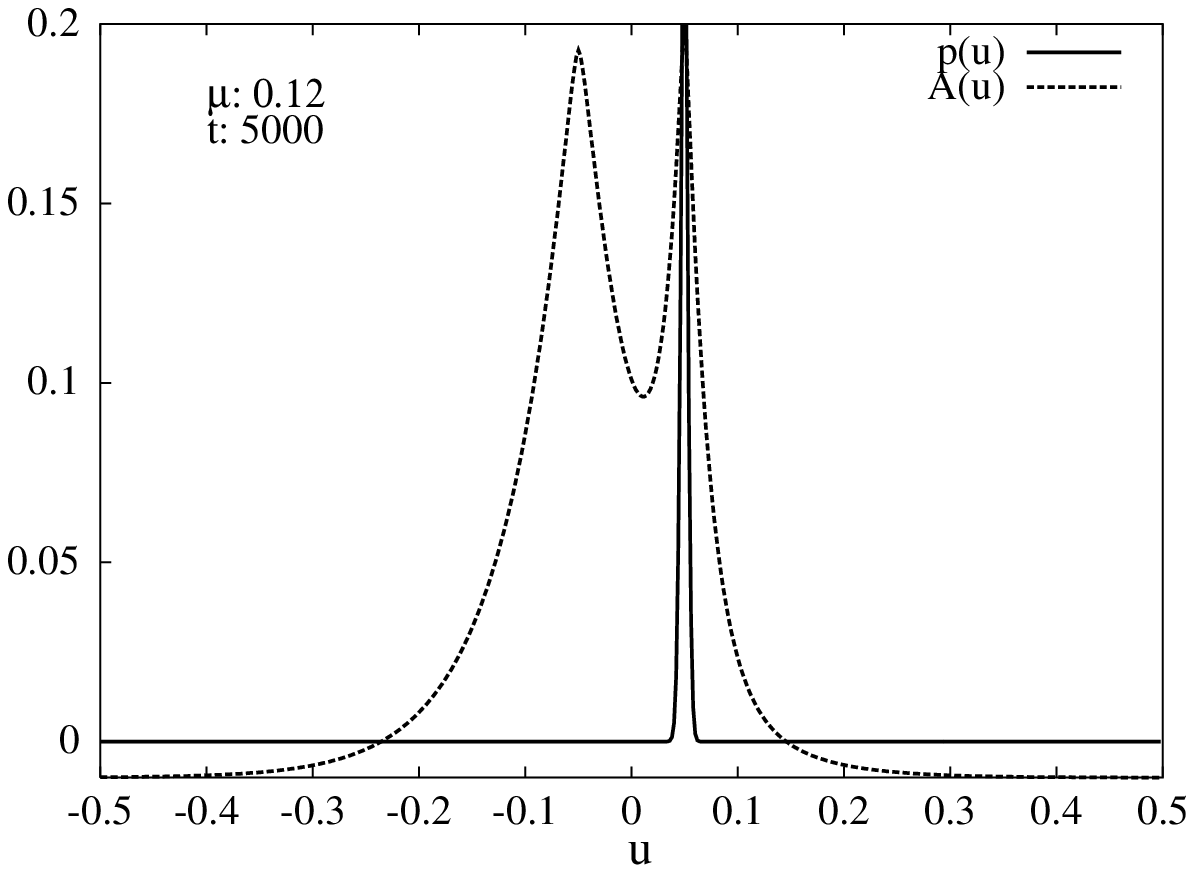}
\includegraphics[width=0.35\columnwidth]{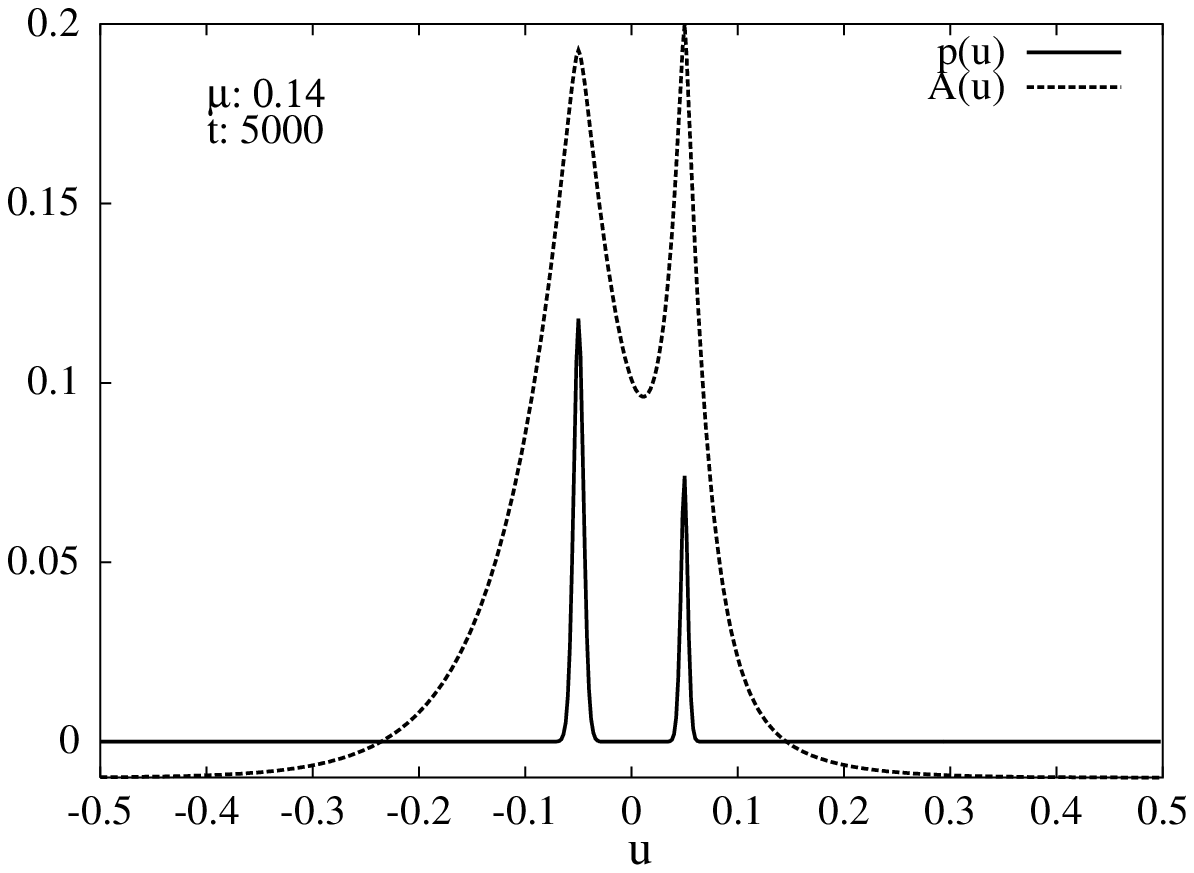}}
\caption{\label{figure:mu-dep} Mutation-induced speciation.
A two peaks static fitness landscape, increasing the mutation rate 
we pass from a single quasi-species population (left, $\mu=0.12$) to the coexistence of
two quasi-species (right, $\mu=0.14$).  
}
\end{figure}

\begin{figure}[Ht]
\centerline{\includegraphics[width=0.35\columnwidth]{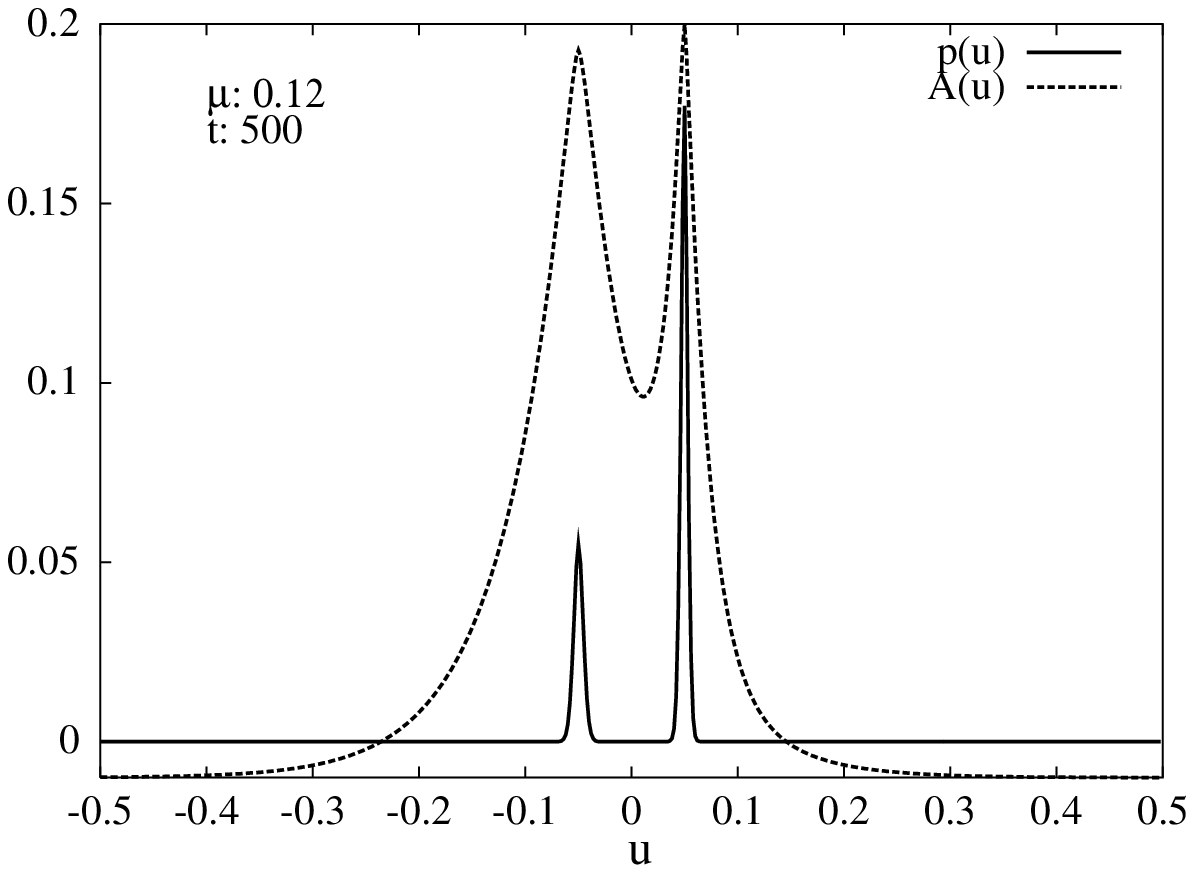}
\includegraphics[width=0.35\columnwidth]{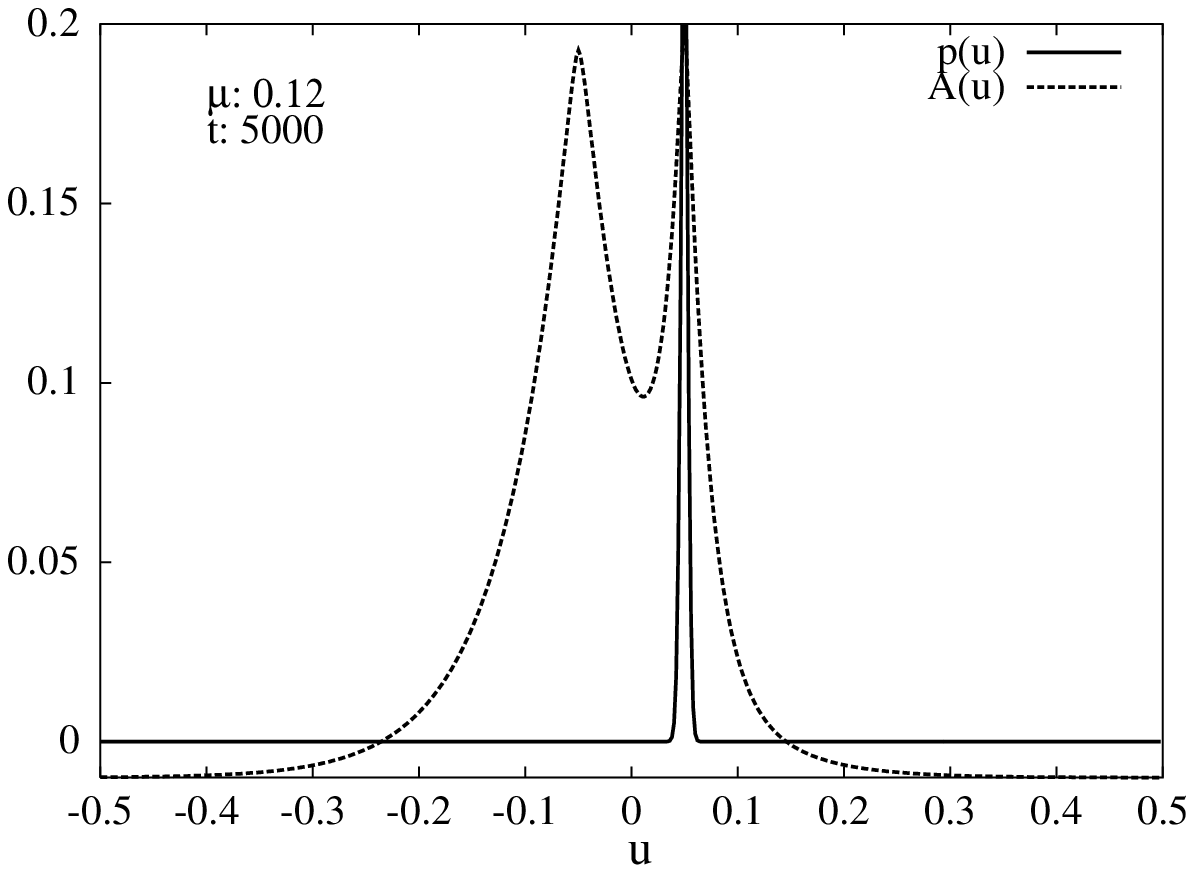}}
\caption{\label{figure:transient} Evolution on a two-peaks
static fitness landscape, after $500$ (left) and $5000$ (right) time steps.
For a transient period of time the two species co-exist, 
but in the asymptotic limit
only the fittest one survives.
}
\end{figure}

Since the average fitness~(\ref{aveA}) of a quasi-species depends on the height $A_0$ and the curvature $a$ of the niche, one can have coexistence of a sharper niche with larger fitness together with a broader niche with lower fitness, as shown in Figure~\ref{figure:mu-dep}. This coexistence depends crucially on the mutation rate $\mu$. If $\mu$ is too small, the quasi-species occupying the broader niche disappears; if the mutation rate is too high the reverse happens. In this case, the difference of fitness establishes the time scale, which can be quite long. In presence of a fluctuating environment, these time scales can be long enough that the extinction due to global competition is not relevant. A transient coexistence is illustrated in Figure~\ref{figure:transient}. One can design a special form of the landscape that allows the coexistence for a finite interval of values of $\mu$, but certainly this is not a generic case. This condition is known in biology under the name of Gause or \emph{competitive exclusion} principle~\cite{Gause}. 
 
We  show in the following that the existence of a degenerate effective fitness is a generic case in the presence of competition, if the two species can co-adapt before going extinct. 

\subsection{Flat landscapes and neutral evolution}
\label{sec:Neutral}

The lineage is the sequence of connections from a son to the father at birth. Going backward in time, from an existing population, we have a tree, that converges to the last universal common ancestor (LUCA). In the opposite time direction, we observe a spanning tree, with lot of branches that go extinct. The lineage of an existing individual is therefore a path in genetic space. Let us draw this path by using the vertical axes for time, and the horizontal axes for the genetic space. Such paths are hardly experimentally observable (with the exception of computer-generated evolutionary populations), and has to be reconstructed from the differences in the existing populations.

The probability of observing such a path, is given by the probability of survival, for the vertical steps, and the probability of mutation for the horizontal jumps. While the latter may often be considered not to depend on the population
and the genome itself\footnote{the inverse of this probability gives the ``molecular clock''}, the survival is in general a function of the population. Only in the case of neutral evolution, one can disregard this factor. Within this assumption, it is possible to compute the divergence time between two genomes, by computing their differences, and reconstruct phylogenetic trees.

\subsection{A simple example of fitness landscape}

The main problem in theoretical evolutionary theory is that of forecasting the consequences of mutations, \emph{i.e.}, the fitness landscape. If a mutation occurs in a coding region, the resulting protein changes. The problem of relating the sequence of a protein to its tri-dimensional shape is still unsolved. Even more difficult is the problem of computing the effects of this modified shape in the chemical network of a cell. Therefore, the only landscape that is considered is the neutral (flat) one.

One of the few cases in which it is possible to guess the consequences of mutations is the case of synonymous substitutions in bacteria~\cite{BagnoliLio}.

The genetic code (correspondence between codons and amino acids) is highly redundant: there are 64 different codons, and only 20 amino acids, plus a stop command that signals the end of the protein chain. This redundancy occurs in two ways: the third codon is often uninfluenced for the binding of the tRNA, and many different tRNA (cognate to different codons) correspond to the same amino acid (synonymous tRNA). Therefore, all these substitutions are neutral with respect to the protein produced.  Synonymous tRNA occurs in different abundances.
\begin{figure}
\centerline{\includegraphics[width=0.7\columnwidth,height=0.5\columnwidth]{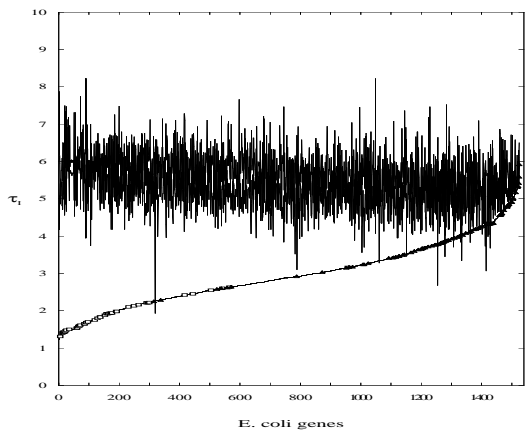}}
\caption{ Mean translation time $\tau _{\text{I}%
  } $ for 1530 {\it Escherichia coli} coding regions calculated using data on
  tRNA abundances published by Ikemura\protect\cite{Ikemura}. The sequences are ordered according to
  the values of $\tau _{\text{I}}$. The lower line shows $\tau _{\text{I}}$
  for the correct reading frame. The upper lines show the value of $\tau _{%
  \text{I}}$ for the +1 and +2 reading frames. The open squares correspond to
  ribosomal genes, the filled triangles correspond to genes carried by
  plasmids.}
  \label{fig:TranslationTime}
\end{figure}

The time needed for a ribosome to attach a new amino acid to the growing chain depends mainly on the waiting time for the right charged tRNA to arrive. This time is proportional to the abundance of tRNAs, and supposing that all of them are always charged, it is proportional to the copies of the tRNAs in the genome.  Therefore, even if synonymous substitutions are neutral for the functionality of proteins, they imply different time intervals for their production.

Bacteria in an abundant medium grow exponentially, and a small difference in the time needed for a duplication can make a big difference in fitness. Proteins are needed in various abundances for duplications: structural ones, like ribosomal proteins, are essential (and consequently they are well optimized and very conserved), while those carried by plasmids, being transmittable among different strains, are presumably not very optimized. This assumption is confirmed by Figure~\ref{fig:TranslationTime}, in which it is reported the translation time $\tau_1$ (computed using the abundances of cognate tRNAs) for various proteins. One can see that the time obtained shifting the frame of +1 or +2 basis, that produces a meaningless protein, is substantially higher for ribosomal protein, and similar to the correct one for proteins carried by plasmids~\cite{Bagnoli:CodonUsage}. 

We can simplify the problem by grouping ``abundant'' tRNAs by the symbol 0, and ``rare'' tRNAs by 1. Disregarding ``intermediate'' tRNAs, we get a sequence $x$ of zeros and ones, and a fitness, assumed monotonic with the replication time $\tau_1$, that has the form 
\[
  A(x) = \exp(\tau_1(x)) = \exp\left(A+B\sum_i x_i\right),
\]
\emph{i.e.}, a monotonic function of the ``magnetization'' of an Ising chain. Here the codons does not interact, and their effect is additive. If one considers that the discharging of a tRNA may slower the translation of cognate codons, one may introduce other couplings among codons~\cite{BagnoliGuasti,Guasti}.

\subsection{Finite populations and random drift}

In finite population, the fluctuations plays a fundamental role. Let us just consider a simple system $\boldsymbol{x}$ composed by $N$ individuals belonging either to variant (or species) A ($x=0$) or B ($x=1$). At each time step an individual $x_i$ is chosen at random, replicates with a probability $p(x_i)$ and substitutes another randomly chosen individual (Moran process~\cite{Moran}). We have a birth-death process of the quantity $n=\sum_i x_i$ with two absorbing states, $n=0$ and $n=N$ corresponding to the extinction of B and A, respectively. It can be shown~\cite{Nowak:EvolutionaryDynamics} that the probability $S(1)$ that a single mutant B will take over the whole population is
\[
 S(1) = \dfrac{1-r}{1-r^N}\quad \mathop{\longrightarrow}_{r\rightarrow 1} \quad \frac{1}{N},
\]
where $r=p(1)/p(0)$ is the selective advantage (or disadvantage) of variant B with respect to A. 
If $\mu \simeq 0$ is the ``mutation probability'' from A to B, the probability of appearance of a mutant in a population of $N$ A's is $\mu N$, and therefore the probability of fixation of a mutant is $Q=\mu N S(1)$, for neutral selection ($r=1$) we get $Q=\mu$, irrespective of the population size~\cite{Kimura}. This result by Kimura provides the \emph{molecular clock} for estimating evolution times from the measure of the distance between two genomes (with the assumption of neutral evolution).

\subsection{Speciation in a flat landscape}

The reasons for the existence of species is quite
controversial~\cite{Maynard:Genetic}.
The definition itself of a species is
not a trivial task. One definition (a) is based simply on the phenotypic
differentiation among the phenotypic clusters, and can be easily applied also to asexual organisms, such as bacteria.
Another definition (b), that applies only to sexual
organism, is based on the inter-breeding possibility. Finally (c), one can
define the species on the basis of the genotypic distance of
individuals, taking into consideration their genealogical
story~\cite{Evolution}.

We shall see in Section~\ref{sec:Sex} a model of how such a segregation, preparatory to speciation, can arise. The role of genetic segregation in maintaining the structure of the  species is particularly clear in the model of Refs.~\cite{DerridaPeliti91,HiggsDerrida91,HiggsDerrida92}. In this model, there is no fitness selection, and one can have asexual reproduction, or sexual (recombination) one with or without a segregation based on genetic distance. Their conclusion  is that species (defined using
the reproductive isolation, definition (b)) can appear in
flat static landscapes provided  with sexual reproduction and
discrimination of mating. In some sense these authors have identified
definitions (b) and (c).
 
In the first model reproduction is asexual: each offspring chooses its parent at random in the previous generation, with mutations.  In this model there are no species, and the distribution fluctuates greatly in time, even for very large populations. The addition of competition, Section~\ref{sec:PhenotypicSpeciation}, may stabilize the distribution. 

In the second model reproduction is recombinant with random mating allowed between any pair of individuals. In this case, the population becomes homogeneous and the genetic distance between pairs of individuals has small fluctuations which vanish in the limit of an infinitely large population. Without segregation (due to competition and/or to genetic incompatibilities) one cannot have the formation of stable, isolated species. 

In the third model reproduction is still recombinant, but instead of random mating, mating only occurs between individuals which are genetically similar to each other. In that case, the population splits spontaneously into species which are in reproductive isolation from one another and one observes a steady state with a continual appearance and extinction of species in the population.

\section{Ageing}\label{sec:Penna}

A simple example of age-dependent (pleyotropic) effects of genes is given by the Penna model~\cite{Penna,Stauffer:SexMoney}. In this model, each individual is characterized by a genome of length $L$, and an age counter. The genome is arranged in such a way that gene $x_i$ is activated at age $i$. There are two alleles, a good one $x_i=0$ and a bad one $x_i=1$. If a given number $m$ of bad genes are activated, the individual dies. 
 
The phenotype $u(x, a)$ depends here of the genotype $x$ and of the age $a$, and can be written as
\[
  u(x, a) = \sum_{i=1}^a x_i,
\]
while the fitness is sharp: 
\[
 A(u) = \begin{cases} 
         1 & \text{if $u<m$,}\\
         0 & \text{otherwise.}
        \end{cases}
\]

This model can explain the shape of Gompertz's  mortality law~\cite{Azbel}. If one imposes that reproduction occurs only after a certain age, one can easily explain the accumulation of bad genes after that age, as exhibited by the catastrophic senescence of many semelparous animals (like the pacific salmon)~\cite{Stauffer:Salmon}. By inserting maternal cares, one can also give a motivation for the appearance of menopause in women~\cite{Sousa}: after a certain age, the fitness may be more increased by stopping giving birth to more sons, with the risk of loosing own life and that of not-yet-independent offspring. 

\section{Dynamic ecosystems}
\label{sec:DynamicalEvolution}

\subsection{Speciation in the phenotypic space}
\label{sec:PhenotypicSpeciation}

We are here referring to the
formation of species in a spatially  homogeneous environment, i.e.\ to
\emph{sympatric} speciation. In this frame of reference, a niche is
 a phenotypic realization of relatively high fitness.
 Species have obviously to do with niches, but one cannot assume that 
 the coexistence of 
species simply reflects the presence of ``pre-existing'' 
niches; on the contrary,
what appears as a niche to a given individual is co-determined by the 
presence of  other individuals (of the same or of different species).
In other words, niches are the product of co-evolution.

In this section we introduce a new factor in our model ecosystem: a
short-range (in phenotypic space) competition among individuals. As
usual, we start the study of its consequences by considering the
evolution in phenotypic space~\cite{Bagnoli:prl,Bagnoli:ecal}. 

We assume that the static fitness $V(u)$, when not flat, is a 
linear decreasing function of the phenotype $u$
except in the vicinity of $u=0$, where it has a quadratic maximum:
\begin{equation}
  V(u) = V_0 + b\left(
    1-\dfrac{u}{r} - \dfrac{1}{1+u/r}
  \right)\label{V}
\end{equation}
so that close to $u=0$ one has 
$V(u) \simeq V_0 -b u^2/r^2$ and for $u\rightarrow \infty$,
$V(u) \simeq V_0 + b(1-u/r)$. 
The master sequence is located at $u=0$.

We have checked numerically that the results are
qualitatively independent on the exact form of the static fitness, providing
that it is a smooth decreasing function. We have introduced this particular form
because it is suitable for analytical computation, but a more classic
Gaussian form can be used.

For the interaction matrix $W$ we have chosen the following  kernel $K$ 
\[
  K\left(\dfrac{u-v}{R}\right) = \exp\left(-\dfrac{1}{\alpha}
  \left|\dfrac{u-v}{R}\right|^\alpha\right).
\]
The parameter $J$  and $\alpha$ control
the intensity and the steepness of the intra-species competition,
respectively. We use a Gaussian ($\alpha=2$) kernel, for the
motivations illustrated in Section~\ref{sec:PhenotypicSpeciation}.    

\begin{figure}[Ht]
\centerline{\includegraphics[width=10cm]{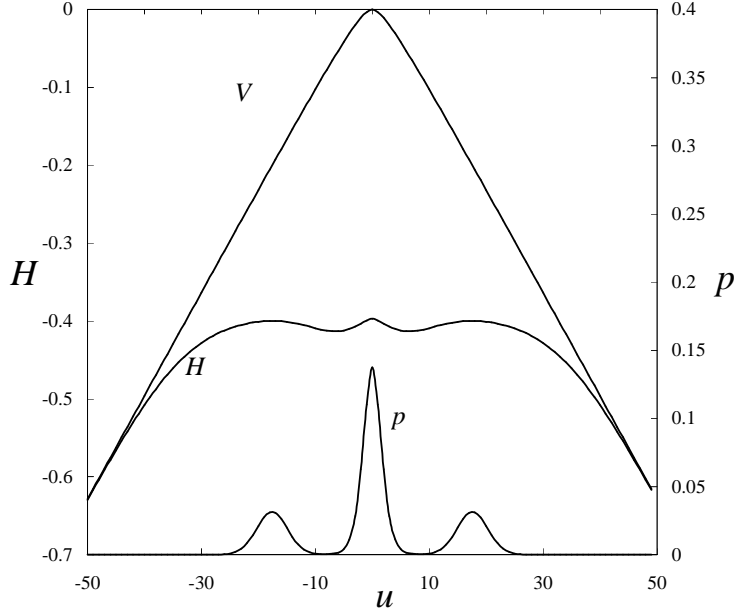}}
\caption{\label{Potential}
Static fitness $V$, effective fitness 
$H$,  and asymptotic distribution $p$ 
numerically computed for the following values of
parameters: $\alpha=2$, $\mu=0.01$, $V_0=1.0$, 
$b=0.04$, $J=0.6$, $R=10$, $r=3$ and $N=100$.
}
\end{figure}

For illustration, we report 
in Figure~\ref{Potential} the numerical solution of
Eq.~(\ref{p}), showing
a possible evolutionary
scenario that leads to the coexistence of three quasi-species. We have
chosen the
smooth static fitness $V(u)$ of Eq.~(\ref{V})
and a Gaussian ($\alpha=2$) competition 
kernel. One can realize that 
the effective fitness  $H$ is almost 
degenerate (here $\mu>0$ and the competition effect extends on
the neighborhood of
the maxima), i.e.\ that the average fitness of all coexisting 
quasi-species is the same. 

We  now derive the conditions for the coexistence of multiple species. 
 We are interested in its asymptotic behavior in the limit
$\mu\rightarrow 0$, which is the case for actual organisms. 
Actually, the mutation mechanism is needed only to define
the genotypic distance and to populate all available niches.
Let us assume that the asymptotic distribution is formed by ${\mathcal L}$ 
quasi-species. Since $\mu\rightarrow 0$ they are approximated by delta
functions $p_k(u)=\gamma_k \delta_{u,u_k}$, $k=0, \dots, {\mathcal L}-1$,
centered at $u_k$.
The weight of each quasi species is $\gamma_k$, i.e.
\[
   \int p_k(u) du = \gamma_k, \qquad\sum_{k=0}^{{\mathcal L}-1} \gamma_k = 1.
\]
The quasi-species are ordered such as  $\gamma_0 \ge
\gamma_1,  \dots, \ge \gamma_{{\mathcal L}-1}$. 

 The evolution equations for the $p_k$ are 
\[
   p_k'(u) = \dfrac{A(u_k)}{\ave A} p_k(u),
\]
where $A(u) = \exp\left(H(u)\right)$ and
\[
  H(u) = 
    V(u) - J\sum_{j=0}^{{\mathcal L}-1} K\left(\dfrac{u - u_j}{R}\right) \gamma_j.
\]

The stability condition of the asymptotic distribution is 
$(A(u_k) - \ave A) p_k(u) = 0$, i.e. either
 $A(y_k) = \ave A = {\rm const}$
(degeneracy of maxima) or $p_k(u)=0$ (all other points). This supports
our assumption of delta functions for the $p_k$. 

The position $u_k$ and the weight $\gamma_k$ of the quasi-species
are given by $A(u_k) = \ave A = {\rm const}$ and 
$\left.{\partial A(u)}/{\partial u}\right|_{u_k} = 0$, or, in terms of the
fitness $H$, by
\[
  V(u_k) - J \sum_{j=0}^{{\mathcal L}-1} K\left(\dfrac{u_k-u_j}{R}\right)
     \gamma_j = \rm{const}
\]
\[
  V'(u_k)  - \dfrac{J}{R}\sum_{j=0}^{{\mathcal L}-1}
  K'\left(\dfrac{u_k-u_j}{R}\right)
   \gamma_j = 0
\]
where the prime in the last equation denotes differentiation with
respect to $u$. 

Let us compute the phase boundary for coexistence of three species for two
kinds of kernels: the exponential ($\alpha=1$)
and the Gaussian ($\alpha=2$) one. The diffusion kernel can be 
derived by a simple
reaction-diffusion model, see Ref.~\cite{Bagnoli:prl}. 

We assume that the static fitness $V(u)$ of Eq.~(\ref{V}). 
Due to the symmetries of the problem, we have the master quasi-species at
$u_0=0$ and, symmetrically,
two satellite quasi-species at $u=\pm u_1$. Neglecting the mutual influence of
the two marginal quasi-species, and considering that $V'(u_0) =
K'(u_0/R)=0$, 
$K'(u_1/R) = -K'(-u_1/R)$, $K(0)=J$ 
and that the three-species threshold is given by $\gamma_0=1$ and $\gamma_1=0$,
 we have 
\[
  \tilde{b}\left(1-\dfrac{\tilde{u}_1}{\tilde{r}}\right) 
        - K(\tilde{u}_1) = -1,  
\]
\[
  \dfrac{\tilde{b}}{\tilde{r}} + K'(\tilde{u}_1) = 0.
\]
where $\tilde{u}=u/R$, $\tilde{r} = r/R$ and $\tilde{b} = b/J$. 
We introduce the parameter $G=\tilde{r}/\tilde{b}=
(J/R)/(b/r)$, that is the ratio of two
quantities, the first one related to the strength of inter-species interactions
($J/R$) and the second to intra-species ones ($b/r$).

In the following we drop the tildes for convenience.
Thus
\[
 r - z - G \exp\left(-\dfrac{z^\alpha}{\alpha}\right) = -G,
\]
\[
 G z^{\alpha-1}\exp\left(-\dfrac{z^\alpha}{\alpha}\right) = 1,
\]

For $\alpha=1$ we have the coexistence condition
\[
 \ln(G) = r -1 + G.
\]
The only parameters that satisfy these equations are $G=1$ and $r=0$,
i.e.\ a
 flat landscape ($b=0$) with infinite range interaction ($R=\infty$). 
Since the coexistence region reduces to a single point,
it is suggested that $\alpha=1$ is a marginal case.
Thus for less steep potentials, such as power law decrease, 
the coexistence condition is supposed not to be
fulfilled.  

For $\alpha=2$ the coexistence condition is given by
\[
  z^2 -(G+r)z + 1 = 0,
\]
\[
  Gz\exp\left(-\dfrac{z^2}{2}\right) = 1.
\]
One can solve numerically this system and obtain the boundary 
$G_c(r)$ for the coexistence. In the limit $r \rightarrow 0$ (almost 
flat static fitness) one has 
\begin{equation}
  G_c(r) \simeq G_c(0) - r \label{Gc}
\end{equation}
with $G_c(0) = 2.216\dots$. 
Thus for $G>G_c(r)$ we have coexistence of three or more quasi-species, while 
for $G<G_c(r)$ only the fittest one survives.

\begin{figure}[Ht]
\centerline{\includegraphics[angle=270,width=10cm]{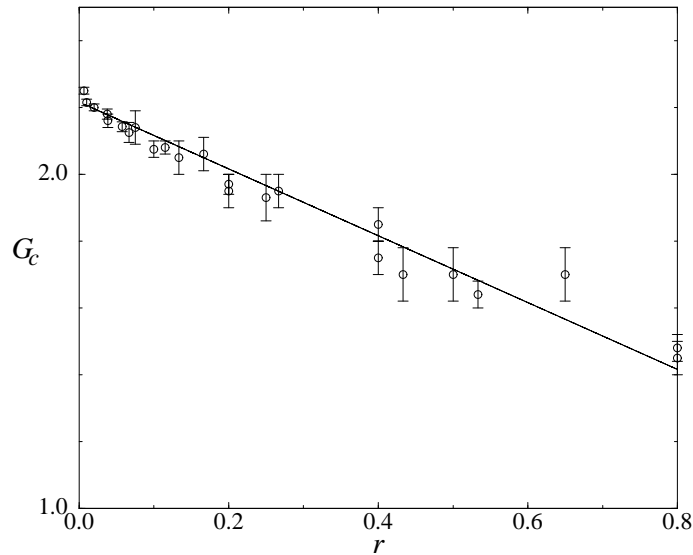}}
\caption{\label{figG}
Three-species coexistence boundary $G_c$ for $\alpha=2$. The continuous 
line represents the analytical 
approximation, Eq.~(\ref{Gc}), the circles are obtained from
numerical simulations. The error bars represent the maximum error (see text for
details).
}
\end{figure}

We have solved numerically  Eq.~(\ref{evol1}) for several
different values of the parameter $G$.
We have considered a discrete phenotypic
 space, with $N$ points, and a simple Euler
algorithm. The results, presented in Figure~\ref{figG}, are not 
strongly affected by the integration step.
The error bars are due to the
discreteness of the changing parameter $G$. 
The boundary of the multi-species phase is well approximated by Eq.~(\ref{Gc});
in particular, we have checked that this 
 boundary does not depend on the mutation rate
$\mu$, at least for $\mu < 0.1$, which can be considered
a very high mutation rate for
real organisms. The most important effect of $\mu$ is the broadening of
quasi-species curves, which can eventually merge as described in
Section~\ref{sec:SmoothLandscape}.

This approximate theory to derive the condition 
of coexistence of multiple quasi-species 
still holds for the hyper-cubic genotypic space.
The different structure of genotypic space does not
change the results in the  limit $\mu \rightarrow 0$.
Moreover, the threshold between one and multiple quasi-species is
defined as the value of parameters for which the satellite
quasi-species vanish. In this case the multiplicity factor of satellite
quasi-species does not influence the competition, and thus we believe
that the threshold $G_c$ of Eq.~(\ref{Gc}) still holds in the
genotypic hyper-cubic space. 

For
a variable
population, the theory still works  
substituting $G$ with  
\begin{equation}
G_a= N G = N \dfrac{J/R}{b/r}, \label{Ga}
\end{equation}
  (for a detailed analysis
see Ref.~\cite{Bagnoli:Review}).

This result is in a good agreement with numerical simulations, 
as shown in the following Section.

\subsection{Speciation and mutational meltdown in the hyper-cubic genotypic space}
\label{sec:Hypercubic}

Let us now study the consequences of evolution in presence of competition 
in the more complex genotypic space. We were not able to obtain
analytical results, so we resort to numerical simulations. Some
details about the computer code we used can be found in Appendix~B. 

In the following we  refer always to rule {\bf (a)}, that allows us to
study both speciation and mutational meltdown. 
Rule {\bf (b)} has a similar behavior for speciation transition, while, of
course, it does not present any mutational meltdown transition.
 
We considered the same static fitness landscape of Eq.~(\ref{V}),  (non-epistatic interactions among genes).

\begin{figure}[Ht]
\centerline{\includegraphics[height=0.8\columnwidth,angle=270]{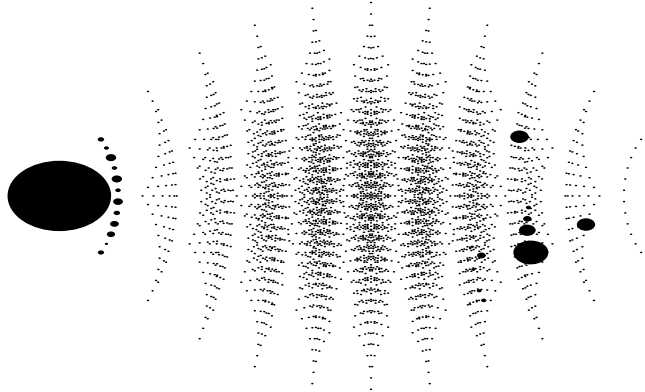}}
\caption{\label{fig:Spot} Easter egg representation of quasi-species in hyper-cubic space for
$L=12$. The smallest points represent placeholder of strains (whose
population is less than $0.02$), only the
larger dots correspond to effectively populated quasi-species; the
area of the
dot is proportional to the square root of population. Parameters:
$\mu=10^{-3}$, $V_0=2$, $b=10^{-2}$, $R=5$, $r=0.5$, $J=0.28$,
$N=10000$, $L=12$.}
\end{figure}

We observe, in good agreement with the analytical approximation 
Eq.~(\ref{Gc}), that
if $G_a$ (Eq.~(\ref{Ga})) is larger than the threshold $G_c$
(Eq.~(\ref{Gc})),
several quasi-species coexist, otherwise only the master sequence
quasi-species survives. 
In Figure~\ref{fig:Spot} a distribution with multiple
quasi-species is shown.

\begin{figure}[Ht]
\centerline{\includegraphics[width=10cm]{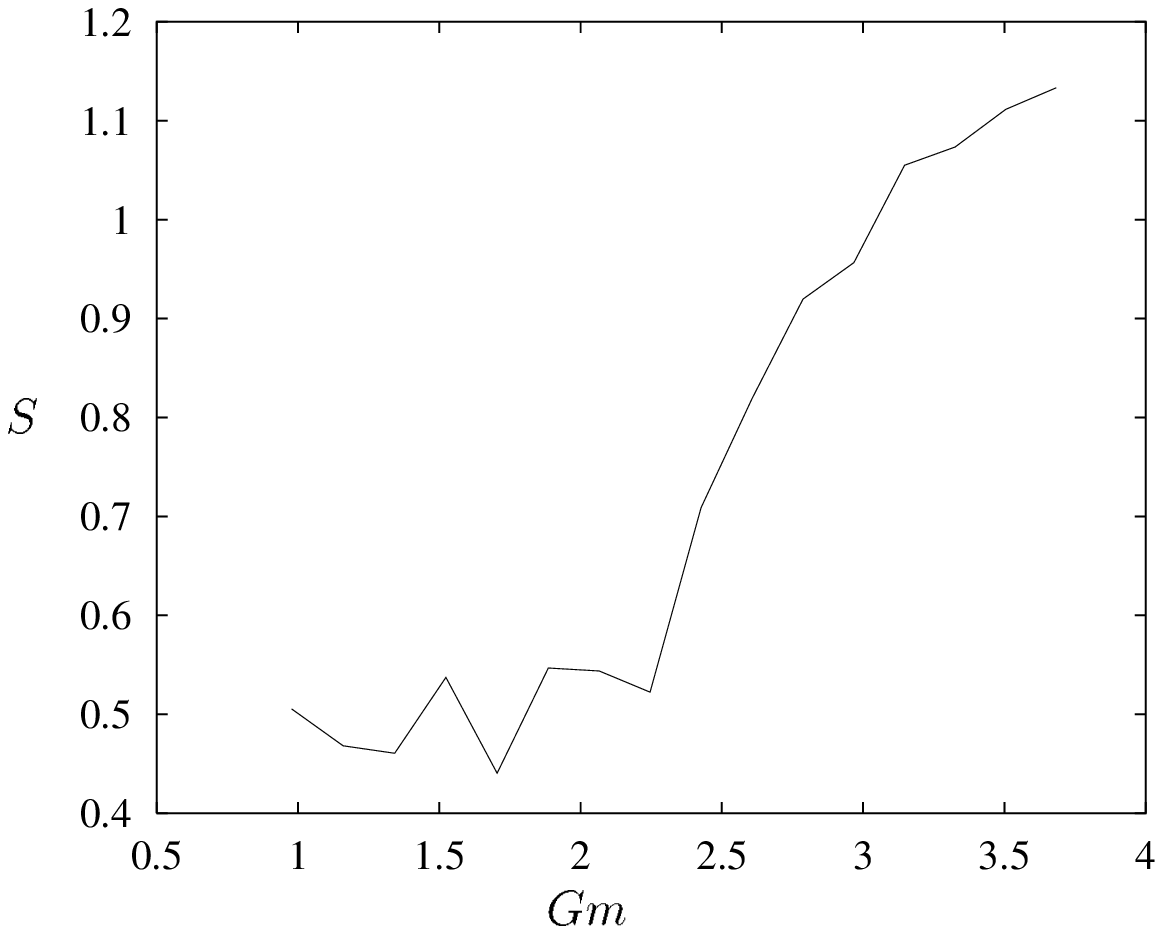}}
\caption{\label{fig:Entropy} The speciation transition characterized
by the entropy $S$ as a function of the control
parameter $G\;m$. Each point is an average over 15 runs.  Same parameters
as in Figure~\protect\ref{fig:Spot}, varying $J$. Errors are of the order of
fluctuations.}
\end{figure}

We can characterize the speciation transition 
by means of the entropy $S$ of the asymptotic phenotypic distribution 
$p(u)$ as function of
$G_a$, 
\[ 
    S = - \sum_u p(u) \ln p(u)
\]
which increases in correspondence of the appearance of multiple
quasi-species (see Figure~\ref{fig:Entropy}).

\begin{figure}[Ht]
\centerline{\includegraphics[width=6.5cm]{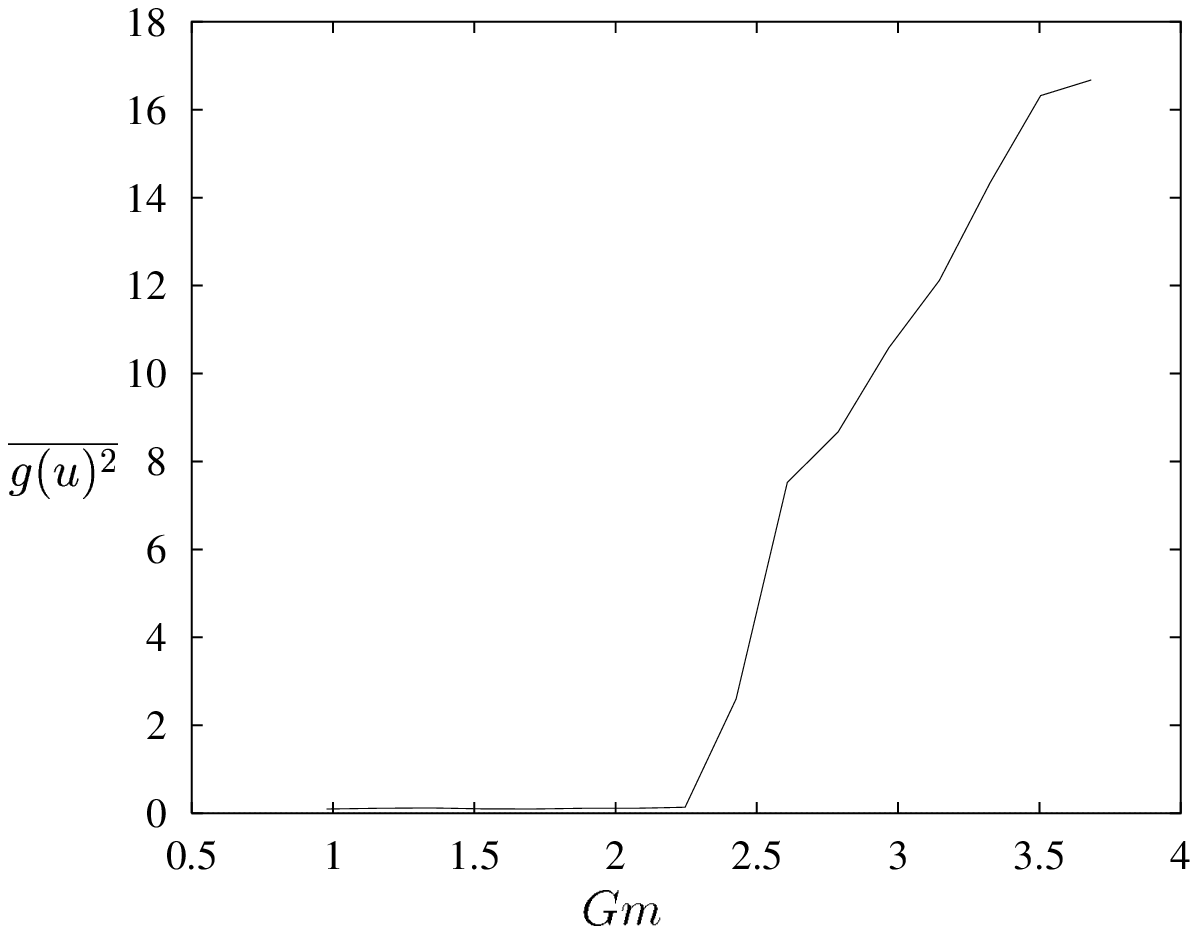}
\includegraphics[width=6.5cm]{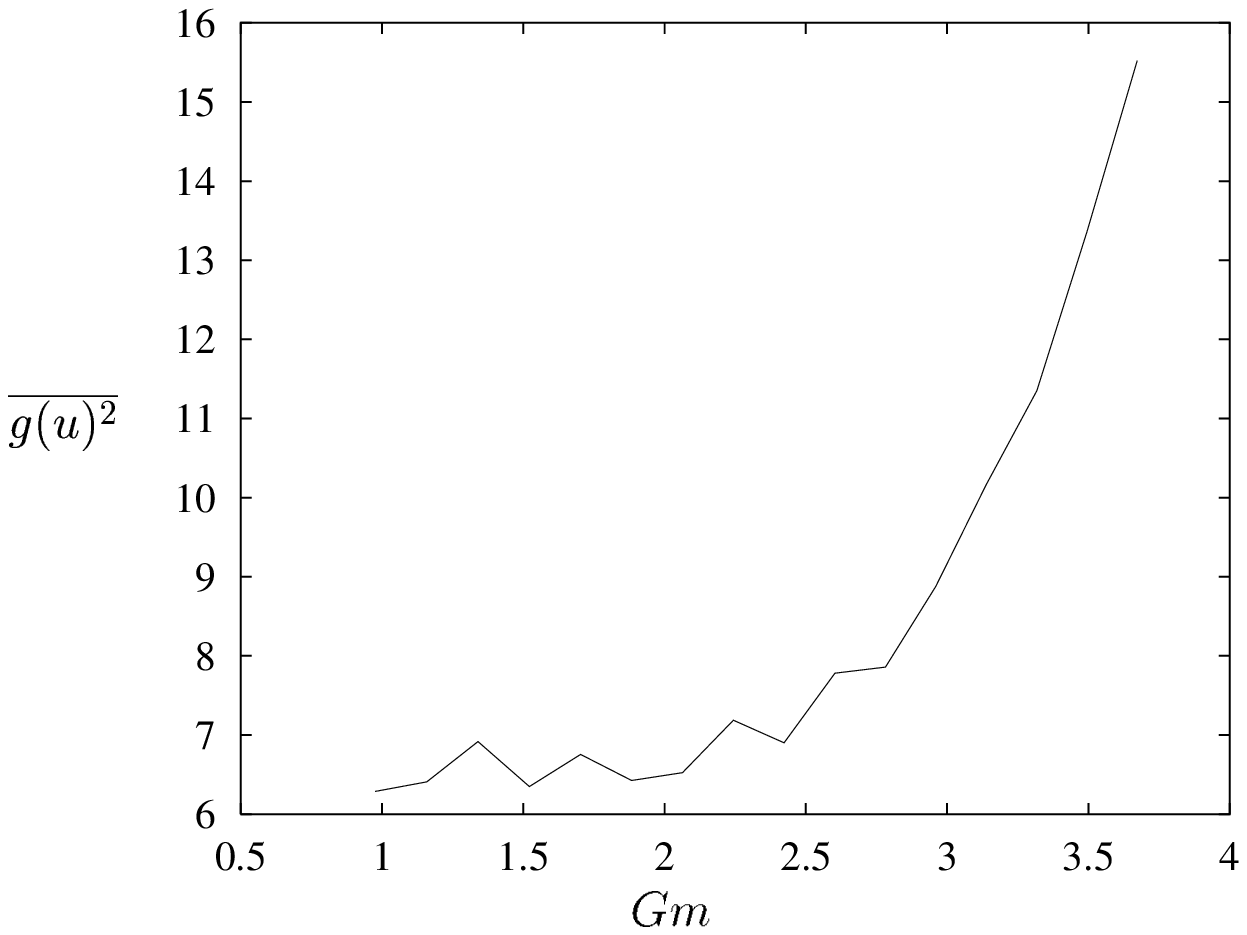}}
\caption{\label{fig:x2} Independence of 
the speciation transition by the mutation rate. 
The transition is characterized by
the average square phenotypic distance $\ave{g(u)^2}$ 
of phenotypic distribution $p(u)$, as a function of the control
parameter $G\;m$. Each point is a single run.  
Same parameters
as in Figure~\protect\ref{fig:Spot}, varying $J$ with $\mu=10^{-3}$
(left) and $\mu=5\;10^{-2}$ (right). }
\end{figure}

We locate the transition at a value $G_a \simeq 2.25$, while 
analytical approximation predicts $G_c(0.1)\simeq 2.116$.
The entropy, however, is quite sensible to
fluctuations of the quasi-species centered at the master sequence
(which embraces the
largest part of distribution), 
and it was necessary to average over several runs in order
to obtain a clear curve; for larger values of $\mu$ it was impossible
to characterize this transition. A quantity which is much less 
sensitive of fluctuations is the
average square phenotypic distance from the master sequence
 $\ave{g(u)^2}$ 
\[
  \ave{g(u)^2}=\sum_u g(u)^2 p(u).
\]
In Figure~\ref{fig:x2} (left)  we characterize the speciation transition by
means of $\ave{g(u)^2}$, and indeed a single run was sufficient,
for $\mu=10^{-3}$. For much higher mutation rates ($\mu=5\; 10^{-2}$) the
transition is less clear, as shown in Figure~\ref{fig:x2} (right),
but one can see that the transition point is substantially
independent of $\mu$, as predicted by the analysis of speciation in the
phenotypic space, Eq.~(\ref{Gc}).  

%
%
%
%

\subsection{Sex}\label{sec:Sex}

By inserting sexual reproduction (recombination) and possibly horizontal transfer of genetic material (conjugation in bacteria, viruses and -- in plants, also tumor-inducing bacteria) we should rather consider a different phylogenetic path for each gene. The history of a given gene (more appropriately, a locus) is in general different from that of another one, still present in the same genome. The survival probability of a gene depends on how good the gene performs in conjunction with the other genes. Even if only ``egoistic'' genes (those that maximize their survival and reproductive probability) passes from a generation to the other, they are nevertheless forced to collaborate. Moreover, genes are identical, in the sense that two individuals or cells may carry the \emph{same} gene. The survival of a gene may thus profit of the death of another individual with the same gene, if this death increases sufficiently the survival and reproductive efficiency of the former. This is the origin of multicellular beings (in which reproduction is reserved to the germ line), of insect colonies, familiar structure and many other evolutionary patterns. More in this subject in Section~\ref{GameTheory}.

Actually, forced cooperation is true only for those genes that may pass to the following generation through the reproduction of the individual. Genes may not follow the fate of the individual by escaping as viruses, or being horizontally transferred (plasmids). Especially in case of viruses, the direct advantage of viruses may correspond to a great damage for the host. If for some mutation the virus content is ``trapped'' inside the host and forced to follow the usual germ line to propagate, that automatically it is forced to collaborate.

In case of sexual reproduction, the evolution may lead to other non-optimal stable patterns, represented by the run-off development of sexual characters. This phenomenon if particularly evident in a model by Dieckmann and Doebeli~\cite{Doebeli}.

In this model, the space plays no role (no spatial niches, sympatric speciation). The fitness in  phenotypic space presents a smooth maximum, so in the absence of other ingredients, the asymptotic distribution is a quasispecies peaked around this maximum. The introduction of a finite-range competition broadens the distribution, but in the absence of assortativity (preference in mating), no speciation occurs~\cite{BagnoliGuardiani}. 

In the model, the assortativity is coded in a portion of the genotype, and is related to a phenotypic character either related to the fitness or acting just like an arbitrary marker. In both cases, speciation may occur (depending on the parameters), even if in the second case the differentiation takes longer (since it is due to genetic drift).

Sex and recombination have important consequences also in the evolution over simple landscapes, like the sharp peak one. In Ref~\cite{SexPhaseTransition} it is shown that while for asexual reproduction the error threshold is driven by the average mutation probability per unit of genome, $\mu$, in recombinant organisms it depends on the total mutation rate $\mu L$, and the transition is near $\mu L=1$. This observation poses a limit to the maximum length of genomes for a given accuracy of replication. 

\subsection{Game theory}\label{GameTheory}

Up to now, we have supposed that there is an instantaneous response to the environment. Considering only binary interactions, the contribution to the score of phenotype $x$ given by an encounter with phenotype $y$,  $W(x|y)$, can schematically be grouped as 
\begin{equation}
\begin{minipage}{0.8\columnwidth}
\begin{center}
\begin{tabular}{cl}
$W(x|y)<0$ \& $W(y|x)<0$: & competition, \\
$W(x|y)>0$ \& $W(y|x)<0$: & predation or parasitism of $y$ on $x$,\\
$W(x|y)<0$ \& $W(y|x)>0$: & predation or parasitism of $x$ on $y$,\\
$W(x|y)>0$ \& $W(y|x)>0$: & cooperation. 
\end{tabular}
\end{center}
\end{minipage}\label{W}
\end{equation}

However, the actual interactions among individuals with memory and strategy are more complex, may depend on the past encounters and on the environment status. 

A more sophisticated modeling follows the ideas of game theory: the genotype of an individual is read as a small program, and the score of an encounter depends on the running of the programs of the participants~\cite{Sigmund}. 

In order to reduce this complexity, let us assume that there are only binary encounters, and that the participants can assume only two \emph{external} status: zero and one, that traditionally are called cooperation or defection. We assume also that the encounter is divided into rounds (hands), and that the number of hands depends on external factors (one could alternatively have a three-letter alphabet like cooperate, defect or escape, and so on). In other words, we consider the participants as programmed automata. 

A generic program therefore says what to play at hand $i$, given the results of previous hands, the identity of the two participants and the internal state (memory). The choice of the output state can be stochastic or deterministic, but in general one has to determine, for each hand $i$, the probability $p(i)$ of outputting 0 (that of outputting 1 is $1-p(i)$), and this can be done using a look-up table that constitutes the genetic information of the individual. 

To make things simpler, let us assume that the choice $i$ does depend only on the past hand, so that a program has to say just how to start (hand 0) and the four probabilities $p(a_x,a_y)$: probability of playing 0 if in the last hand player $x$ played $a_x$ and player $y$ played $a_y$. More complex strategies can be used~\cite{Axelrod}.

The score of the hand for player $x$ is given by the payoff matrix $Q(a_x|a_y)$. The score for player $y$ is the transpose of $Q'$ if the game is symmetric.  The interesting situation is the \emph{prisoner's dilemma}: $Q(1|0) > Q((0|0) > Q(1|1) > Q(0|1)$.~\footnote{The asymptotic state of the iterated Prisoner's dilemma game is not given by a ``simple'' optimization procedure, see Ref.~\cite{Peliti:Games}} For a single-shot play, standard game theory applies. It can be shown that it is evolutionary convenient for both player to defect, since mutual cooperation ($Q(0|0)$) gives less payoff that exploitation ($Q(1|0)$); which from the point of view of the opposite player (transpose) gives the minimal payoff. The game is interesting if  cooperation gives the higher \emph{average} payoff ($2Q(0|0)>Q(0|1)+Q(1|0)$). 
  
Axelrod~\cite{Axelrod} noted that if the the game is iterated, cooperation can arise: rule with memory, like TIT-FOR-TAT (TFT: cooperate at first hand, and then copy your opponent's last hand) may stabilize the mutual cooperation state, and resist invasion by other simpler rules (like ALL-DEFECT -- ALLD, always play 1). Other more complex rules may win against TFT (and possibly loose against ALLD).  

Nowak~\cite{Nowak} gives at least five ways for which mutual cooperation can arise in an evolutionary environment: 
\begin{enumerate}
 \item Kin selection. Cooperation is preferred because individuals share a large fraction of the genome, so a gene for cooperation that is common in the population  shares the average payoff, even in a single-shot game.
 \item Direct reciprocity. This is just Axelrod's iterated game. If the expected length of the game is large enough, it is worth to try to cooperate for gaining the higher share. 
 \item Reputation. Information about past games can be useful in deciding if own opponent is inclined towards cooperation or towards exploitation. 
 \item Network reciprocity. In this case one considers the spatial structure of the group, represented by a graph with local connectivity $k$. A player plays against all neighbors, and gather the related score. If the advantage in cooperating in a cluster of  $k$ neighbors is greater than the gain I would have by switching to defect, the strategy is stable. 
 \item Group selection. In this case, the system is assumed to be split in groups, each of which can become extinct, or give origin to another one. A group can be invaded by defectors, but in this case it will succumb in competition with all-cooperator groups, that have larger payoffs.  
\end{enumerate}

\subsection{Effective treatment of strategies}

The analysis of strategies with memory is quite complex, due to the large number of variables. As shown by Nowak~\cite{Nowak}, it is often possible to study the evolutionary competition among strategies by using an effective payoff matrix $W$, that specifies the average gain of a strategy against the others, given the parameters of the problem. In this way the evolution of the system is just given by the interaction weights of Eq.~\eqref{W}, and the analysis is quite simpler. 

For the competition between two strategies A and B with a parameter $p$, three conditions can be found~\cite{Nowak}, according with the position of the saddle point $p_c$ that separates the basins of all-A ($p=0$) and all-B ($p=1$)distributions:
\begin{enumerate}
\item If $p_c=0$, only the distribution all-B is present. This generally corresponds to all-defectors.
\item If $p_c$ is near zero, the strategy all-A is \emph{evolutionary stable}: in an infinite population, a single (or small but dispersed) mutant B cannot invade, while a bunch of B (in neighboring sites) can.
\item If $p_c>1/2$, the A startegy is \emph{risk-dominant}: by mixing at random the population, in a larger number of sites A is dominant.
\item If $p_c> 2/3$ then the A strategy is \emph{advantageous}. The analysis of finite population of size $N$ shows that the fixation probability of a mutant by random drift for flat or weak selection is $1/N$. For $p_c> 2/3$, the fixation probability of a cooperator mutant in a finite population of defectors is greater than that given by random drift.    
\end{enumerate}

\subsection{Self-organization of ecosystems}\label{SOC}

The final goal of a theoretical evolutionary theory is to explain the macro-patterns of evolution (speciation, extinctions, emergence of new characteristics like cooperation, formation of complex ecosystems) as the result of self-organization of individuals~\cite{Maynard:multispecies}. The main idea is that at the molecular level evolution is either neutral, or lethal. The latter corresponds to unviable phenotypes, for instances, mutations that lead to proteins that do not fold correctly. Within this assumption, the fitness space for a single ``replicator'' (an ancient bio-molecule) is like a slice of a Swiss cheese. We assume that there is one big connected component. By considering only the paths on this component, the origin of life should correspond to one largely connected ``hub'', where entropy tends to concentrate random walks (induced to mutations).

However, competition (red queen), due to the exhaustion of resources, soon changes the fitness of different genotypes. In particular, the location corresponding to the origin of life becomes the one where competition is maximal, due to the large connectivity. The competition induces speciation, so that a hypothetical ``movie'' of the ancient lineages will show a diverging phylogenetic tree, in which the locations corresponding to ancestors become minima of fitness (extinction) after a speciation event. As we have seen, competition also promotes the grouping of phenotypes and the formation of ``species'', even in an asexual world.

In experimental and simulated evolutionary processes, mutants soon arises trying to exploit others by ``stealing'' already formed products, like for instance the capsid of viruses. This is the equivalent of a predator/prey (or parasite/host) relationship. Predators induce complex networks of relationship, for which even distant (in phenotypic space) ``species'' coevolve synchronously. When such an intricate ``ecological'' network has established, it may happen that the extinction of a species may affect many others, triggering an avalanche of extinctions (mass extinctions). 

The opposite point of view is assuming that chance and incidents dominate the evolution. This could be a perfect motivation for punctuated equilibrium~\cite{Gould} (intermittent bursts of activity followed by long quiescent periods): random catastrophic non-biological events (collision of asteroids, changes in sun activity, etc.) suddenly alter the equilibrium on earth, triggering mass extinction and the subsequent rearrangements. These events certainly happens, the problem is that of establishing their importance. 

It is quite difficult to verify the first scenario computationally. A good starting point is the \emph{food web}~\cite{FoodWeb}. We can reformulate the model in our language as follows. A species $i$ is defined by a set of $L$ phenotypic traits $i_1, \dots , i_k$, chosen among K possibilities. The relationship between two species is given by the match of these characteristics, according with an antisymmetric $K\times K$ matrix $\boldsymbol{M}$. The score $W(i|j)$ accumulated by species $i$ after an encounter with species $j$ is given by
\[
 W(i|j) = \frac{1}{L} \sum_{n=1}^k\sum_{m=1}^k  M_{i_n,j_m}.
\]
A species accumulates the score by computing all possible encounters. A monotonic function of the score determines the growth (or decreasing) of a species. When the system has reached a stationary state, the size of a species determines its probability of survival (\emph{i.e.}, the survival probability is determined by the accumulated score). When a species disappears, another one is chosen for cloning, with mutations in the phenotypic traits.
  
A special ``species'' 0 represents the source of energy or food, for instance solar light flux. This special species is only ``predated'' and it is not affected by selection. 

This model is able to reproduce a feature observed in real ecosystems: in spite of many coexisting species, the number of tropic level is extremely low, and increases only logarithmically whith the number of species.

The model can be further complicated by adding spatial structure and migration, ageing, etc.~\cite{StaufferChowdhury}. 

Simpler, phenomenological models may be useful in exploring how coherent phenomena like mass extinction may arise from the  self-organization of ecosystems. Most of these models consider species (or niches) fixed, disregarding mutation. 
Possible, the simplest one is the Bak-Sneppen (BS) model~\cite{BakSneppen}. In the BS model, species are simply assigned a location on a graph, representing the ecological relation between two species (in the simplest case, the graph is a regular one-dimensional lattice with nearest-neighbor links). A species is assigned a random number, interpreted as its relative abundance.\footnote{In the original paper this number is termed ``fitness'', but in an almost-stable ecosystem all individuals should have the same fitness.} At each time step, the smallest species is removed, and its ecological place is taken by another one, with a random consistence. With this change, the size of species connected to the replaced one are also randomly changed.
The smallest species at the following time step may be one of those recently changed, or an unrelated one. The first case is interpreted as an \emph{evolutionary avalanche} (mass extinction). 

This \emph{extremal dynamics} (choice of the smallest species) is able to self-organize the ecosystem so that the size distribution of species is not trivial (a critical value appears so that no smaller species are present), and the distribution of avalanche size follows a power-law, similar (but with a different exponent) to that observed in paleontological data by Raup~\cite{Raup}. This model is able to illustrate how self-organization and punctuated equilibrium can arise internally in a dynamical system with long-range coupling (due to the extremal dynamics). 

The Bak-Sneppen model can be though as an extremal simplification of a food web, which in turn can be considered a simplification of the microscopic dynamics (individual-based) of a modeled ecosystem. 

\section{Conclusions}

We have shown some aspects of a theoretical approach to self-organization in evolutionary population dynamics. The ideal goals would be that of showing how macro-evolutionary patterns may arise from a simplified individual-based dynamics. However, evolutionary systems tend to develop highly correlated structure, so that is it difficult to operate the scale separation typical of simple physical system (say, gases). One can develop simple models of many aspects, in order to test the robustness of many hypotheses, but a comprehensive approach is still missing.


\end{document}